\documentclass{elsarticle}
\usepackage{hyperref}
\usepackage[utf8]{inputenc}
\usepackage{natbib}
\usepackage{graphicx}
\usepackage{latexsym}
\usepackage{amssymb}
\usepackage{amsmath}
\usepackage{amsfonts}
\usepackage{gensymb}
\usepackage{bm}
\usepackage{braket}
\usepackage{color}
\usepackage{multirow}
\usepackage{tikz}
\usetikzlibrary{arrows}
\usepackage[normalem]{ulem} 

\newcommand{\rcite}[1]{Ref.~\cite{#1}}
\newcommand{\eqnref}[1]{Eq.~(\ref{#1})}
\newcommand{\figref}[1]{Fig.~\ref{#1}}
\newcommand{\sfigref}[2]{Fig.~\hyperref[#1]{\ref{#1}(#2)}}

\newcommand{\secref}[1]{Sec.~\ref{#1}}
\newcommand{\appref}[1]{\ref{#1}}

\newcommand{\vx}{\mathbf{x}}
\newcommand{\tT}{{\tilde{t}}}
\newcommand{\tX}{{\tilde{x}}}
\newcommand{\tY}{{\tilde{y}}}
\newcommand{\tZ}{{\tilde{z}}}
\newcommand{\R}{x^\mu}
\newcommand{\tR}{{\tilde{x}^\mu}}

\newcommand{\dd}{\mathrm{d}}
\newcommand{\ii}{\mathrm{i}}

\hypersetup{
  colorlinks = true,
  urlcolor = black,
  pdfauthor = {Wilbur Shirley, Kevin Slagle, Xie Chen},
  pdftitle = {Shirley et al. - 2018 - Fractional excitations in foliated fracton phases}
}

\biboptions{numbers,sort&compress} 

\begin{document}

\begin{frontmatter}

\title{Fractional excitations in foliated fracton phases}
\author[caltechaddress]{Wilbur Shirley}
\author[torontoaddress,caltechaddress]{Kevin Slagle}
\author[caltechaddress]{Xie Chen}

\address[caltechaddress]{Department of Physics and Institute for Quantum Information and Matter, \\ California Institute of Technology, Pasadena, California 91125, USA}
\address[torontoaddress]{Department of Physics, University of Toronto, Toronto, Ontario M5S 1A7, Canada}

\begin{abstract}
Fractional excitations in fracton models exhibit novel features not present in conventional topological phases: their mobility is constrained, there are an infinitude of types,
and they bear an exotic sense of `braiding'. Hence, they require a new framework for proper characterization. Based on our definition of foliated fracton phases in which equivalence between models includes the possibility of adding layers of gapped 2D states, we propose to characterize fractional excitations in these phases up to the addition of quasiparticles with 2D mobility. That is, two quasiparticles differing by a set of quasiparticles that move along 2D planes are considered to be equivalent; likewise, `braiding' statistics are measured in a way that is insensitive to the attachment of 2D quasiparticles. The fractional excitation types and statistics defined in this way provide a universal characterization of the underlying foliated fracton order which can subsequently be used to establish phase relations. We demonstrate as an example the equivalence between the X-cube model and the semionic X-cube model both in terms of fractional excitations and through an exact mapping.
\end{abstract}

\end{frontmatter}

{\hypersetup{linkcolor=black} 
\tableofcontents}

\section{Introduction}
\label{sec:intro}

Gapped topological phases are characterized by their fractional excitations and the universal braiding statistics amongst them. For example, the $\nu=1/3$ fractional quantum Hall state contains $e/3$ fractional charges; exchanging two fractional charges results in a phase factor of $\pi/3$.\cite{WilczekStatistics,ArovasStatistics} In two-dimensional gapped topological phases, possible sets of fractional excitations and their fusion rules and braiding statistics are captured by the mathematical framework of unitary modular tensor categories.\cite{KitaevAnyons} In three spatial dimensions (3D), there are loop-like excitations in addition to point-like excitations. For example, a discrete gauge theory in 3D contains both point-like gauge charges and loop-like gauge fluxes; the exchange of gauge charges, the braiding of a gauge charge around a gauge flux, and the three-loop braiding of (linked) gauge fluxes \cite{Wang2014a,Jiang2014,Wang2015} give rise to universal statistics. In general, it is expected that the set of fractional excitations together with their exchange and braiding statistics provide a complete characterization of the underlying topological order. In other words, if two gapped systems have the same fractional excitations, then they belong to the same topological phase and can be deformed into one another without undergoing a phase transition.

The recent discovery of \textit{fracton} models \cite{ChamonModel,HaahCode,YoshidaFractal,Sagar16,VijayFracton,FractonRev,3manifolds,Slagle17Lattices,CageNet,VijayNonabelian,MaLayers,ChamonModel2,HsiehPartons,HalaszSpinChains,VijayCL,Slagle2spin,FractalSPT,YouSondhiTwisted,SongTwistedFracton,Petrova_Regnault_2017,Slagle17QFT,BulmashFractal} introduces new possibilities \cite{FractonEntanglement,DevakulCorrelationFunc,KubicaYoshidaUngauging,WilliamsonUngauging,Schmitz2018,YouSSPT,Bravyi11,PremGlassy,ShiEntropy,HermeleEntropy,BernevigEntropy,PinchPoint,DevakulWilliamson} and at the same time poses new challenges to this means of characterization.
It was found that in these 3D gapped lattice models, point-like excitations have restricted motion (whereas point-like excitations of topological phases can move freely throughout the entire 3D space). Some excitations, which we will refer to as \textit{lineons}, can only move along a line; others, which we call \textit{planons}, can only move within a plane; the so-called \textit{fractons} are fully immobile as individual particles -- however they may move in coordination as the corners of an expanding or shrinking rectangle or tetrahedron. The excitations are \textit{fractional} in the sense that they cannot be individually created or destroyed by local operations. (We do not assume any global symmetry in these models and are not using the word `fractional' in the sense that the excitations carry fractional symmetry representations such as fractional charge or spin).

It is not obvious how to properly describe these excitations. Point-like excitations in conventional topological phases are grouped into superselection sectors: two excitations belong to the same superselection sector if they can be mapped into each other via local operations.\cite{KitaevAnyons} If we utilize the same notion to describe fractional excitations in fracton phases, the number of superselection sectors is unbounded and grows exponentially with the total system size. Moreover, due to the restricted mobility of the excitations, it is not clear what constitutes a universal quasiparticle `braiding' process. Given such difficulties, it is not clear how to use fractional excitations to characterize and compare the non-trivial orders in different fracton models.

In this paper, we introduce a way of characterizing fractional excitations and their statistics for a sub-class of fracton phases -- the `foliated fracton phases' which we defined in Ref. \cite{3manifolds} and Ref. \cite{FractonEntanglement}. We observed that a large class of fracton models contain a foliation structure -- the system size can be increased by adding two dimensional topological layers and smoothly fusing them in -- such that their non-trivial properties can in large part be attributed to these underlying layers.\footnote{Gapless $U(1)$ fracton models \cite{PretkoU1,electromagnetismPretko,Rasmussen2016,Xu2006,PretkoTheta,PretkoDuality,PretkoGravity,GromovElasticity,MaPretko18,BulmashHiggs,MaHiggs,PaiFractonicLines} and type-II fracton models (for which excitations are created at corners of fractal operators) \cite{HaahCode,YoshidaFractal} are not foliated fracton phases, and will not be considered in this work.} Such properties include the sub-extensive scaling of the logarithm of the ground state degeneracy with system size (linear in the length of the system), and a sub-leading correction to the area law term in the ground state entanglement entropy of a region that scales linearly with the diameter of the region. To unmask the intrinsic 3D nature of the order in these models, we consider models that differ by gapped 2D layers to be equivalent. That is, we define two gapped fracton models to be in the same foliated fracton phase if they can be smoothly deformed into one another (without closing the energy gap to excited states) upon the addition of gapped 2D layers. This definition subsequently points to the natural way to properly describe fractional excitations in foliated fracton phases -- by \textit{modding out} the planons.

More specifically, we generalize the notion of superselection sectors to that of \textit{quotient superselection sectors} (QSS) so that two point-like excitations are considered as equivalent not only if they can be related by local operations, but also if they can be related by attaching a planon (i.e. an excitation that can only move in two dimensions). Under this generalized notion of equivalence, the number of sectors becomes finite, which greatly simplifies the counting. Correspondingly, when we subsequently define quasiparticle statistics using interferometric detection, we only consider processes which are indifferent to the attachment of planons.

This way of describing fractional excitations provides a powerful tool for comparing foliated fracton order in different models. In particular, we show that the X-cube model \cite{Sagar16} and the semionic X-cube model \cite{MaLayers} have the same fractional excitations and statistics according to this definition, despite the fact that their statistics appear very different prior to taking the quotient. This suggests that these two models may have the same foliated fracton order and indeed we present an exact mapping from one to another that involves the addition of 2D topological layers to each model.

This paper is organized as follows: in section~\ref{sec:Xc}, we briefly review the X-cube model, in particular its fractional excitations and foliation structure. Section~\ref{sec:QSS} defines the quotient superselection sectors (QSS) and the subsequent section~\ref{sec:statistics} discusses a a way of characterizing their statistics using interferometric detection. Both sections use the X-cube model as an example to explain the idea. In section~\ref{sec:examples}, several other fracton models are studied, including a novel anisotropic lineon model in section~\ref{sec:anisotropic}. Their fractional excitation content is found to belong to several classes, as summarized in the table in the concluding section~\ref{sec:discussion}. The explicit mapping between the semionic X-cube model to the X-cube model is given in section~\ref{sec:mapping}. In section~\ref{sec:loop}, we briefly extend the discussion to encompass loop excitations of 3D topological orders.

\section{The X-cube model}

\label{sec:Xc}

\begin{figure}[htbp]
\centering
\includegraphics[width=0.8\textwidth]{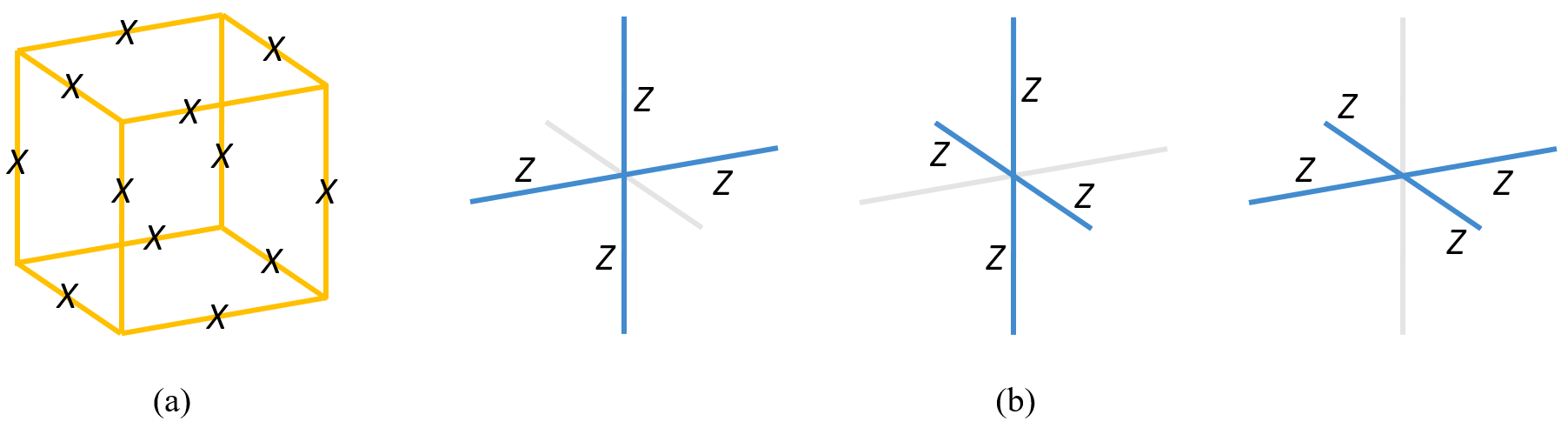}
\caption{(a) Cube and (b) cross terms of the X-cube Hamiltonian.}
\label{fig:ZCH}
\end{figure}

In this section we briefly review the X-cube model, its hierarchy of subdimensional fractional excitations, and the RG transformation for the model which utilizes 2D toric code layers as resource states. As originally discussed in \rcite{Sagar16}, the model is defined on a cubic lattice with a qubit degree of freedom placed on each lattice link. The Hamiltonian is a frustration-free sum of mutually commuting operators (shown in \figref{fig:ZCH}):
\begin{equation}
    H=-\sum_v\left(A^{xy}_v+A^{xz}_v+A^{yz}_v\right)-\sum_c B_c,
    \label{eqn:XcubeH}
\end{equation}
where the first sum is overall vertices $v$ of the lattice, and the second sum is over all elementary cubes $c$. The vertex term $A^{xy}_v$ is equal to a product of Pauli $Z$ operators over the four links emanating adjacent to $v$ within the $xy$ plane (and likewise for $A^{xz}_v$ and $A^{yz}_v$). Conversely, the cube term $B_c$ is equal to a product of Pauli $X$ operators over the twelve edges of the cube $c$. The ground state wavefunction under open boundary conditions may be written as
\begin{equation}
    \ket{\psi_0}=\prod_c(1+B_c)\ket{0}
\end{equation}
where $\ket{0}$ is the simultaneous +1 eigenstate of the Pauli $Z$ operators on all links. It can be helpful to conceptualize this wavefunction as a condensate of extended objects with rectangular prism geometry.

\begin{figure}[htbp]
\centering
\includegraphics[width=0.75\textwidth]{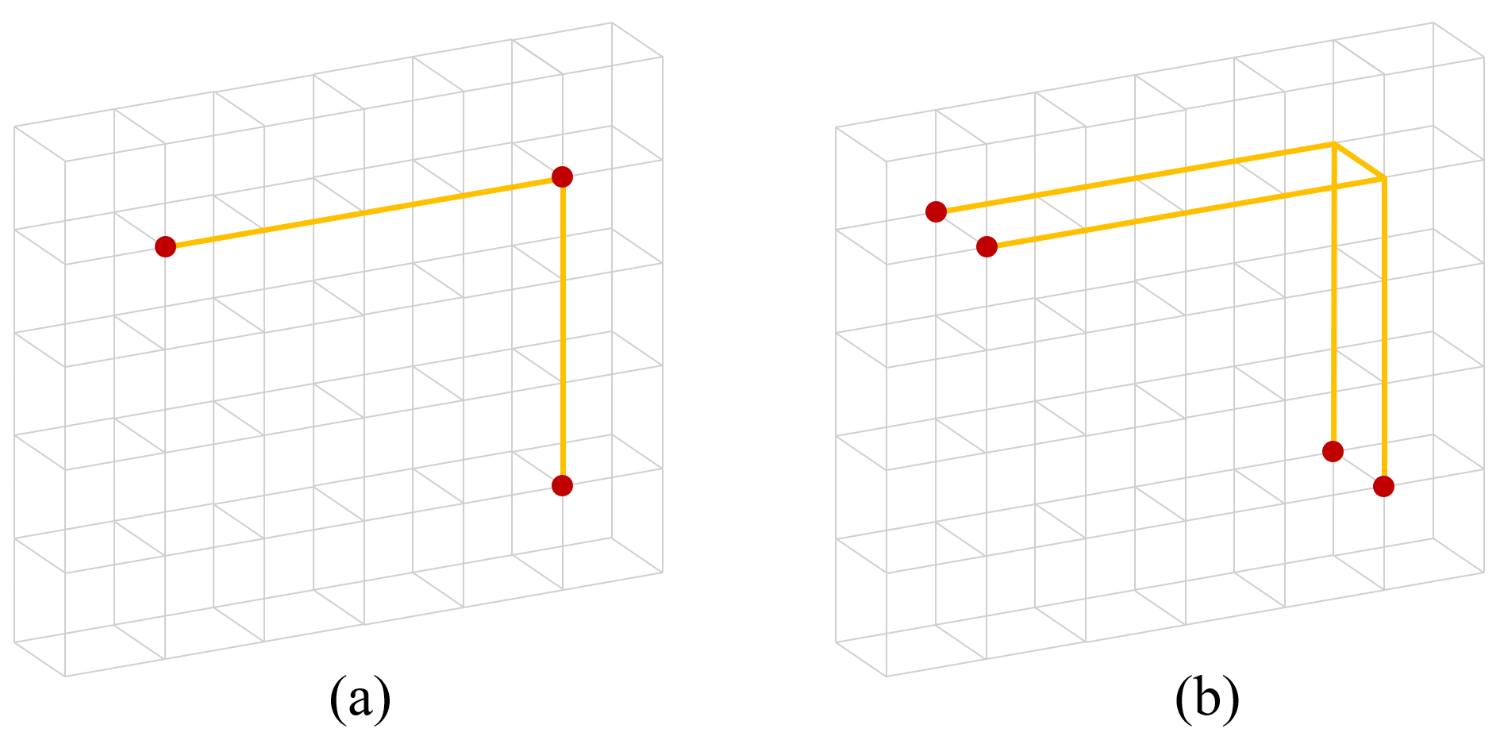}
\caption{(a) A rigid string operators in the X-cube model. Lineons, represented as red dots, are created at the endpoints and corner. (b) A flexible string operator. Lineon dipoles, which are free to move in a 2D plane, are created at the endpoints.}
\label{fig:XCstrings}
\end{figure}

\begin{figure}[htbp]
\centering
\includegraphics[width=0.8\textwidth]{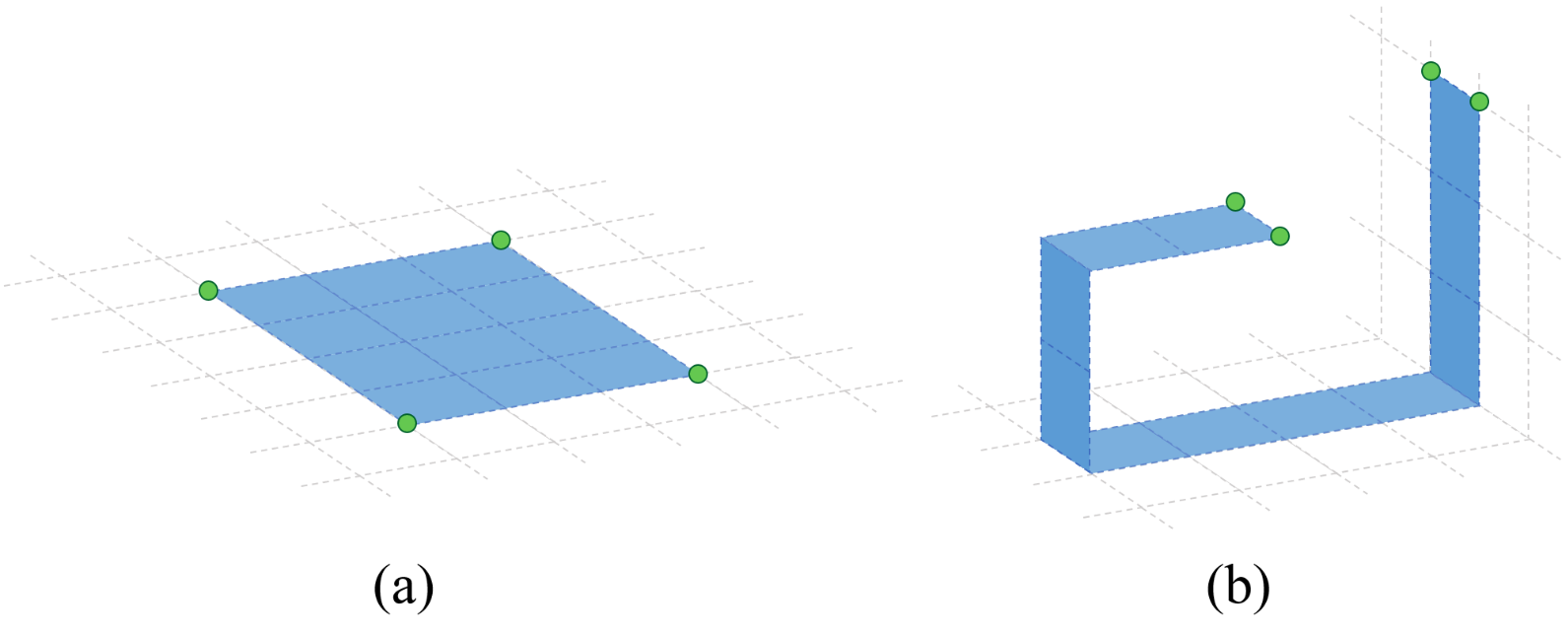}
\caption{(a) A membrane operator. It resides on the plaquettes of the dual lattice, whose edges are depicted as dashed lines. Fractons, represented as green dots, are created at the corners. b) A ribbon operator, which is a type of membrane operator. The fracton dipoles created at its endpoints are free to move in a 2D plane.}
\label{fig:XCmembranes}
\end{figure}

The fractional excitations of the model can be naturally grouped into `electric' and `magnetic' sectors, whose quasiparticles are violations of the cross and cube terms respectively. The electric sector contains three types of lineons (1D particles), which are created at the ends of open \textit{rigid} string operators and move in the $x$, $y$, or $z$ direction (as shown in \sfigref{fig:XCstrings}{a}). These objects obey a triple fusion rule, in which three lineons moving in different directions may collectively annihilate to the vacuum if they meet at a point. Moreover, pairs of adjacent lineons may be viewed as dipolar objects which are themselves fractional planon excitations (2D particles). For example, a pair of lineons mobile in the $x$ direction and separated in the $z$ direction may move within the $xy$ plane via the action of \textit{flexible} string operators (see \sfigref{fig:XCstrings}{b}). On the other hand, the magnetic sector hosts fracton excitations which occur at the corners of open membrane operators. These membrane operators are most naturally thought of in the dual lattice picture in which qubits are attached to elementary plaquettes of the lattice, which are grouped together to form membranes (as shown in \sfigref{fig:XCmembranes}{a}). Pairs of fractons created at adjacent corners of a membrane operator may be viewed as dipolar planons in their own right, which become mobile within a plane via the action of thin membrane operators which we will call \textit{ribbon} operators (see \sfigref{fig:XCmembranes}{b}). These ribbon operators along with the flexible string operators can be thought of as 2D string operators which create planons out of the vacuum at their endpoints.

\begin{figure}[htbp]
\centering
\includegraphics[width=0.8\textwidth]{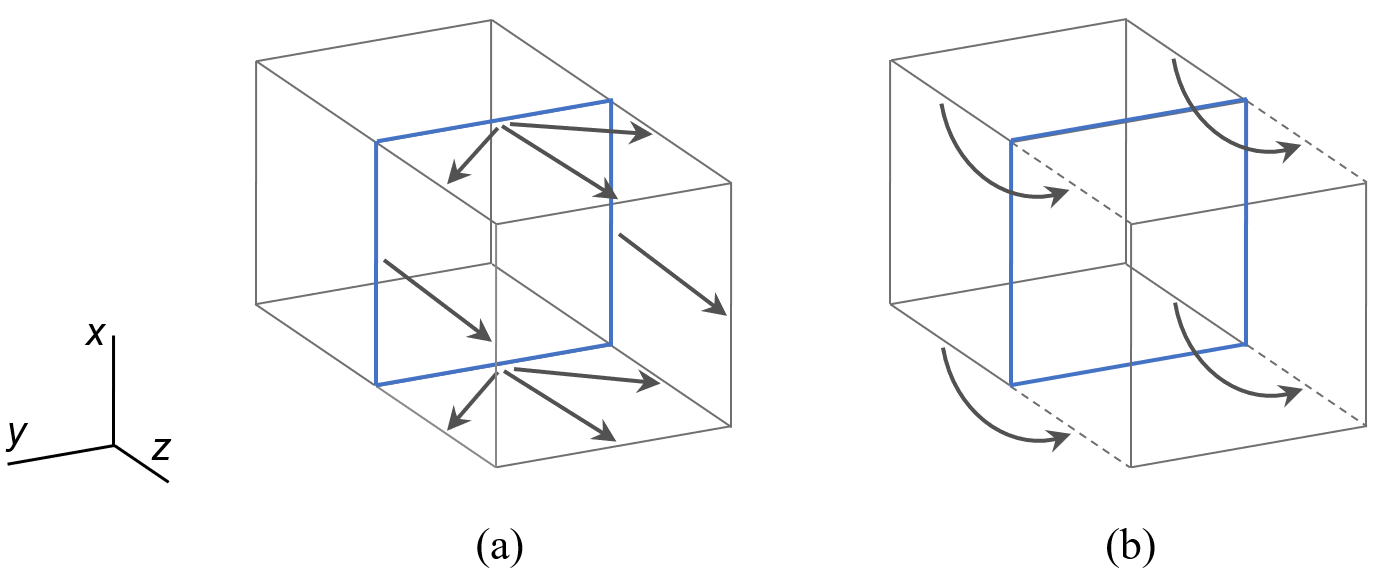}
\caption{A graphical representation of the unitary operators (a) $S_1$ and (b) $S_2$. In this figure only a single unit cell is depicted, although $S_1$ and $S_2$ act uniformly along an $xy$ plane. The finite depth quantum circuit $S=S_1S_2$ disentangles the blue $xy$ layer from the bulk X-cube system. The qubits represented by dashed edges in (b) are decoupled ancilla qubits stabilized by $H_0$ of \eqnref{eqn:conjugation}.}
\label{fig:XCRG}
\end{figure}

The X-cube model has vanishing correlation length, and is actually a fixed point model under a renormalization group (RG) procedure which refines or coarsens the underlying lattice by sewing and un-sewing toric code layers into the system via quantum circuits of finite depth.\cite{3manifolds} The elementary transformation disentangles a single toric code layer from an $L_x\times L_y\times L_z$ size X-cube model to yield a reduced X-cube model on a lattice of dimensions $L_x\times L_y\times (L_z-1)$. This is realized as a finite depth quantum circuit $S$ which satisfies
\begin{equation}
    SH_\mathrm{XC} S^\dagger \cong H'_\mathrm{XC}+H_\mathrm{TC}+H_0,
    \label{eqn:conjugation}
\end{equation}
where $H_\mathrm{XC}$ is the original X-cube Hamiltonian, $H_\mathrm{XC}'$ is the reduced X-cube Hamiltonian, $H_\mathrm{TC}$ is the Hamiltonian of the decoupled toric code layer, and $H_0$ is a trivial Hamiltonian corresponding to ancillary product state degrees of freedom. Here the relation $H\cong H'$ indicates that $H$ and $H'$ have coinciding ground spaces and thus correspond to the same phase of matter. The unitary operator $S$ can be written as the composition $S=S_1S_2$. Here $S_1$ and $S_2$ are commuting tensor products of controlled-NOT 2-qubit gates and are depicted graphically in \figref{fig:XCRG}.

\section{Quotient superselection sectors}
\label{sec:QSS}

\subsection{Review: superselection sectors}

Before defining the notion of \textit{quotient superselection sectors}, we will begin by reviewing the notion of ordinary \textit{superselection sectors}, which correspond to the elementary quasiparticle types of a topological phase. First, let us carve a small, ball-shaped region $\mathcal{R}$ out of a three-dimensional gapped bulk. Suppose that the medium is of infinite spatial extent, and consider the set of all excited states $\ket{\psi_s}$ that are locally indistinguishable from the ground state outside of $\mathcal{R}$, but may contain excitations within $\mathcal{R}$. An ordinary \textit{superselection sector} is a universality class of such states which are related to one another via local unitary operators. To be precise, two normalized states $\ket{\psi_p}$ and $\ket{\psi_q}$ are said to belong to the same superselection sector if there exists a local unitary operator $U$ with support in $\mathcal{R}$ such that $\ket{\psi_p}=U\ket{\psi_q}$. The superselection sector may be subsequently viewed as the subspace spanned by all such equivalent excited states.

Actually, because the system has been posited to have infinite spatial extent, this heuristic discussion does not have a solid footing since there are ambiguities when comparing wavefunctions of infinite extent. However, it can be made rigorous by imagining that we take a finite macroscopic sample $\mathcal{M}$ of the system surrounding $\mathcal{R}$, and map wavefunctions into the space of density matrices on the subsystem $\mathcal{M}$. An arbitrary wavefunction $\ket{\psi}$ corresponds to the reduced density matrix $\rho=\mathrm{tr}_{\overline{\mathcal{M}}} \ket{\psi}\bra{\psi}$, where the degrees of freedom outside $\mathcal{M}$ have been traced out. Two density matrices $\rho_p$ and $\rho_q$ are then considered equivalent if there is a unitary $U$ such that $\rho_p=U\rho_q U^\dagger$. The use of density matrices is implicit in the definitions that follow; however we will omit mention of them as to do otherwise would obfuscate the physical intuition, and in all cases it is a straightforward task to make the definitions rigorous by incorporating their use. 

The \textit{vacuum} sector consists of states containing only local excitations, whereas the non-trivial sectors correspond to fractional excitations of the medium, which cannot be annihilated via local processes. For conventional 3D topological orders, there are a finite number of superselection sectors, corresponding to the point-like topological charges of the phase which are created at the endpoints of open Wilson strings. For example, for a 3+1D discrete gauge theory based on a finite group $G$, the superselection sectors (for a ball-shaped region) correspond to irreducible representations of $G$. Conversely, for foliated fracton phases, the fundamental constraints on quasiparticle mobility give rise to an exponential growth of the number of superselection sectors as the diameter of the region $\mathcal{R}$ is increased, corresponding to an infinite number of fractional excitation types. However, as we will see, by `modding out' the fractional excitations that correspond to anyonic quasiparticles of the underlying foliation layers, i.e. the \textit{planon} sectors of the phase, the resulting \textit{quotient superselection sectors} are finite in number and independent of the size of $\mathcal{R}$.

\subsection{Definition: Quotient superselection sectors}

The RG picture of foliated fracton phases, in which layers of 2D topological orders can be systematically disentangled from the rest of the system, can be used to make the intuition mentioned above precise. For simplicity, consider a state $\ket{\psi_\rho}$ containing only a single planon, labelled $\rho$, in region $\mathcal{R}$, whose plane of mobility is denoted by $\mathcal{P}$. The planon can be `disentangled' from the rest of the system via an RG transformation, i.e. a finite depth quantum circuit $V$ such that $V\ket{\psi_\rho}=\ket{\psi_0'}\otimes\ket{\psi^{2\mathrm{D}}_\alpha}$. Here $\ket{\psi_0'}$ is the ground state of a modified system with inhomogeneities in the vicinity of $\mathcal{P}$, and $\ket{\psi^\mathrm{2D}_\alpha}$ is an excited state of a 2D topologically ordered phase living in plane $\mathcal{P}$, containing an anyonic excitation $\alpha$ in the region $\mathcal{R}\cap\mathcal{P}$.\footnote{The local unitary $V$, viewed as a quantum circuit, has a minimum depth which scales with the spatial extent of $\rho$. For instance, in the case of a dipolar planon composed of two fractons, the depth scales as the distance between the fractons.} The planon $\rho$ can thus be thought of as a fractional excitation belonging to layer $\mathcal{P}$ of the underlying foliation structure.

\begin{figure}[htbp]
\centering
    \includegraphics[width=.975\textwidth]{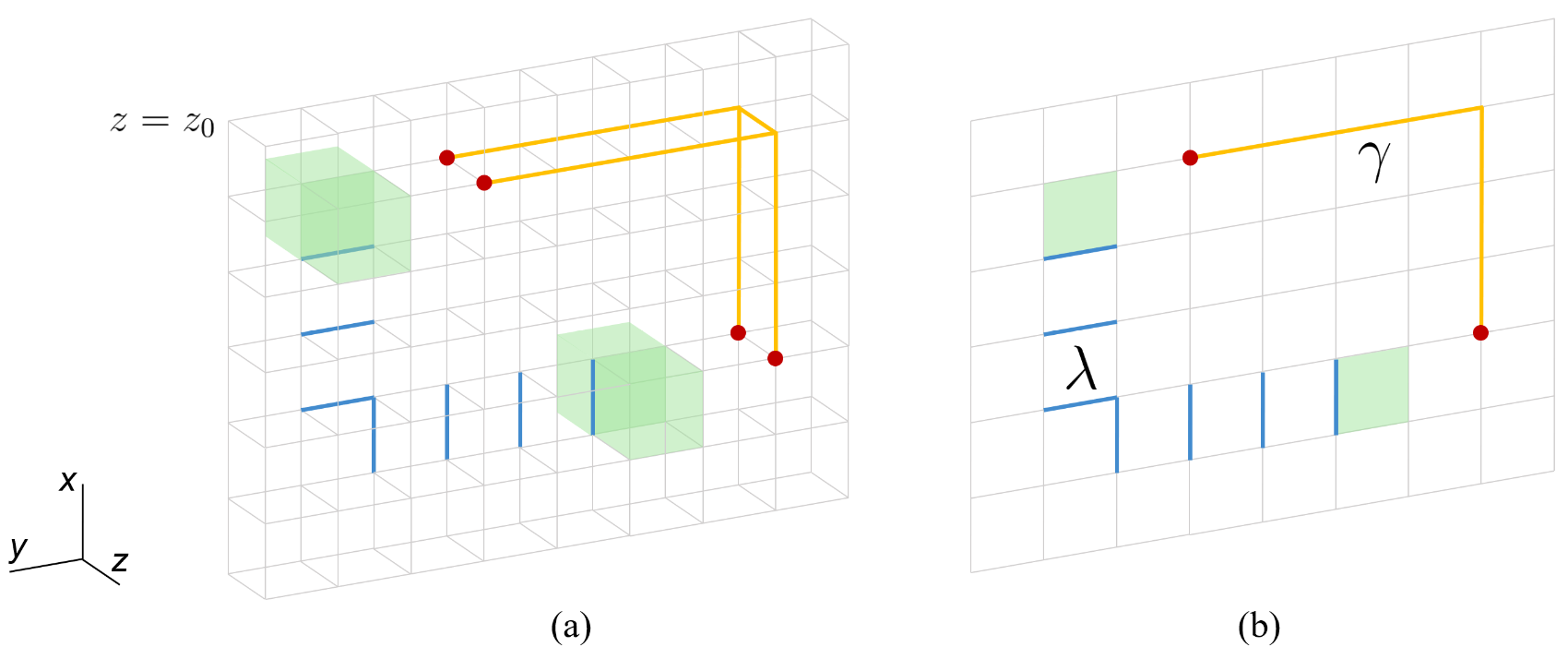}
\caption{(a) A flexible string operator $W_\varepsilon(\gamma)$ and a ribbon operator $W_\mu(\lambda)$ of the X-cube model, which are mapped under conjugation by the finite-depth circuit $S$ to (b) electric and magnetic string operators $W_e(\gamma)$ and $W_m(\lambda)$ acting on a decoupled toric code layer lying along the $z=z_0$ plane, which is the back plane pictured in (a). These operator are respectively defined as tensor products of Pauli $X$ operators over the yellow edges and Pauli $Z$ operators over the blue edges. $\gamma$ and $\lambda$ are paths on the direct and dual lattices respectively of the $z=z_0$ plane. The red dots represent X-cube lineons in (a) and $\mathbb{Z}_2$ charges in (b). Conversely, the shaded green cubes in (a) represent fractons, whereas the green squares in (b) represent $\mathbb{Z}_2$ fluxes.}
\label{fig:XCplanons}
\end{figure}

It is instructive to understand how the disentangling of planons can be achieved in the X-cube model. As discussed, there are two types of planons in the X-cube model: fracton dipoles, and lineon dipoles. Consider a path $\gamma$, with endpoints, lying along the direct lattice edges in the $z=z_0$ plane. Denote by $W_\varepsilon(\gamma)$ a flexible string operator lying alongside $\gamma$ adjacent to the $z=z_0$ plane, for example as shown in \figref{fig:XCplanons}. $W_\varepsilon(\gamma)$ creates lineon dipoles at its endpoints, which are elementary in the sense that they are separated by a single lattice spacing in the $z$ direction. Likewise, consider a ribbon $\lambda$ composed of dual lattice plaquettes which are dual to $x$ and $y$ links in the $z=z_0$ plane, and let $W_\mu(\lambda)$ denote the membrane operator corresponding to $\lambda$, which creates elementary fracton dipoles (i.e. pairs of fractons separated by a single lattice spacing) at its endpoints (see \figref{fig:XCplanons}). Now consider the action of the operator $S$, introduced in the discussion of the X-cube RG transformation in the previous section, which disentangles a toric code layer along $z=z_0$ from the rest of the system. It can be seen that
\begin{align}
    SW_\varepsilon(\gamma)S^\dagger&=W_e(\gamma)
    \label{eqn:strings}\\
    SW_\mu(\lambda)S^\dagger&=W_m(\lambda)
\end{align}
where $W_e(\gamma)$ is an open electric string operator along the path $\gamma$, residing in the disentangled toric code layer along the $z=z_0$ plane, and likewise $W_m(\lambda)$ is an open magnetic string operator in the toric code layer which lies along $\lambda$ (the dual lattice plaquettes comprising $\lambda$ become dual lattice links when restricted to the planar square lattice). $W_e(\gamma)$ creates $\mathbb{Z}_2$ charges at the endpoints of $\gamma$ whereas $W_m(\lambda)$ creates $\mathbb{Z}_2$ fluxes at the endpoints of $\lambda$. From \eqnref{eqn:strings} it follows that $S\ket{\psi_\varepsilon}=\ket{\psi_0'}\otimes\ket{\psi^\mathrm{TC}_e}$ where $\ket{\psi_0'}$ is the ground state of the reduced X-cube Hamiltonian $H_\mathrm{XC}'$, and $\ket{\psi_\varepsilon}$ and $\ket{\psi_e^\mathrm{TC}}$ are excited states of the (original) X-cube and toric code Hamiltonians respectively containing an elementary lineon dipole and a $\mathbb{Z}_2$ gauge charge. Similarly, $S\ket{\psi_\mu}=\ket{\psi_0'}\otimes\ket{\psi^\mathrm{TC}_m}$. The above discussion addresses elementary dipoles of lineons and fractons. For dipolar planons which a larger spatial extent, similar disentangling circuits can be constructed; however, the depth of the circuit scales linearly with the length of the dipole.

With this motivation in mind, we define two (normalized) excited states $\ket{\psi_p}$ and $\ket{\psi_q}$ to belong to the same \textit{quotient superselection sector} (QSS) if there exists a unitary operator $U$ with support in $\mathcal{R}$ and a finite depth quantum circuit $V$ that satisfy
\begin{align}
    U\ket{\psi_p'}&=\ket{\psi_q'} \nonumber\\
    V\ket{\psi_p}&=\ket{\psi_p'}\otimes
        \ket{\psi^{2\mathrm{D}}_{\alpha_1}}\otimes\cdots
        \otimes\ket{\psi^{2\mathrm{D}}_{\alpha_n}} \\
    V\ket{\psi_q}&=\ket{\psi_q'}\otimes
        \ket{\psi^{2\mathrm{D}}_{\beta_1}}\otimes\cdots
        \otimes\ket{\psi^{2\mathrm{D}}_{\beta_n}}, \nonumber
\end{align}
where $\ket{\psi_p'}$ and $\ket{\psi_q'}$ are modified excited states, and $\ket{\psi^\mathrm{2D}_{\alpha_i}}$ and $\ket{\psi^\mathrm{2D}_{\beta_i}}$ are 2D topologically ordered states living along plane $\mathcal{P}_i$ and respectively harboring (possibly vacuous) anyonic excitations $\alpha_i$ and $\beta_i$ in the region $\mathcal{R}\cap\mathcal{P}_i$. The operator $V$ naturally decomposes into a product, $V=\prod_i V_i$, where $V_i$ are operators that subsequently disentangle 2D layers along $\mathcal{P}_i$. According to this definition, it immediately follows that superselection sectors containing only planons belong to the vacuum quotient sector, since the planons will can be mapped into a $\ket{\psi^{2\mathrm{D}}}$ state, as exemplified in \figref{fig:XCplanons}. In this sense, the planon sectors are factored out, and the resulting quotient sectors may be viewed as fractional quasiparticle species \textit{modulo} anyons of the underlying foliation layers. Factoring out the planons in this manner consistently results in a finite set $\mathcal{A}$ of QSS. If a quasiparticle state belongs to a particular quotient sector $a\in\mathcal{A}$, let us say that such an excitation carries \textit{quotient charge} $a$. States within a given sector may be viewed as belonging to a Hilbert space $\mathcal{H}_a$. 

In foliated fracton models, the above definition is equivalent to a more transparent formulation which more closely parallels the definition of ordinary superselection sectors. A \textit{planon creation operator} $W_\rho$ is a unitary operator that has planar support and creates a planon $\rho$ at its endpoint in region $\mathcal{R}$, and extends to spatial infinity at the other end. Then, two states $\ket{\psi_p}$ and $\ket{\psi_q}$ are defined to represent the same quotient superselection sector if there exist planon creation operators $W_{\rho_j}$ ($j=1,\ldots,m$) and a local unitary operator $U$ such that
\begin{equation}
\left(U\prod_j W_{\rho_j}\right)\ket{\psi_p}=\ket{\psi_q}.    
\end{equation}
In other words, an equivalence relation on excited states is imposed which affords the freedom to arbitrarily create and annihilate planons, in addition to local excitations, within $\mathcal{R}$.

\begin{figure}[htbp]
\centering
\includegraphics[width=0.85\textwidth]{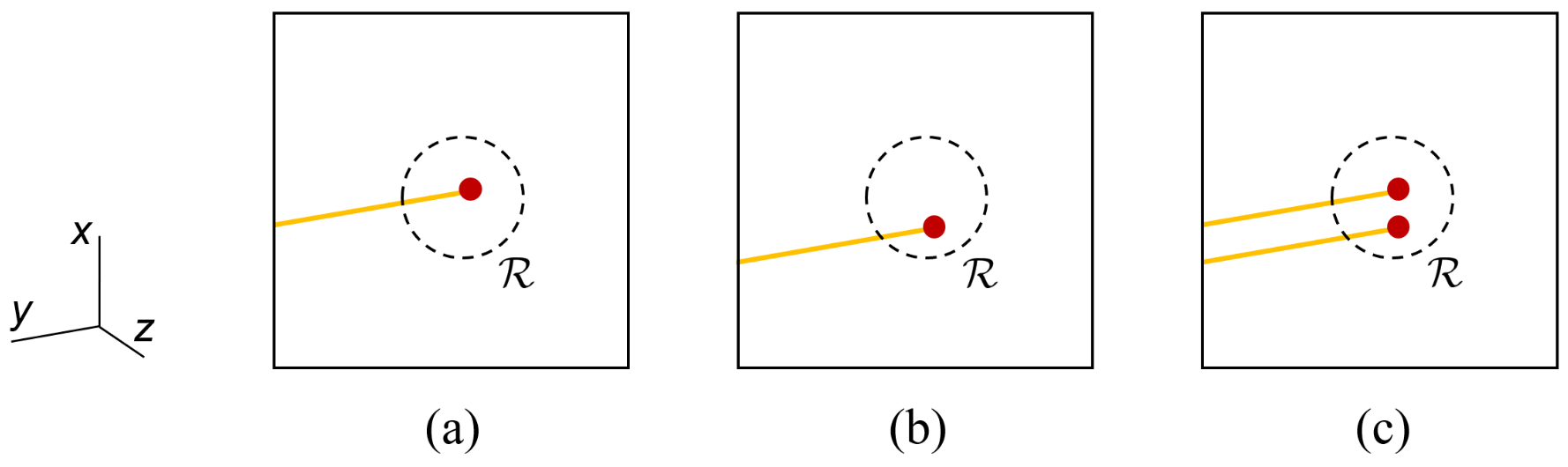}
\caption{(a, b) Representative states of the $\ell_y$ quotient superselection sector of the X-cube model. These states are the result of acting on the ground state with the yellow rigid string operators, which create lineons at their endpoints (red dots) in $\mathcal{R}$. (c) A state containing a planon free to move in a $yz$ plane. The flexible string operator that creates this planon is an effective hopping operator between the lineon states in (a) and (b).}
\label{fig:XCquotient}
\end{figure}

\subsection{Example: X-cube model}

\begin{figure}[htbp]
\centering
\includegraphics[width=\textwidth]{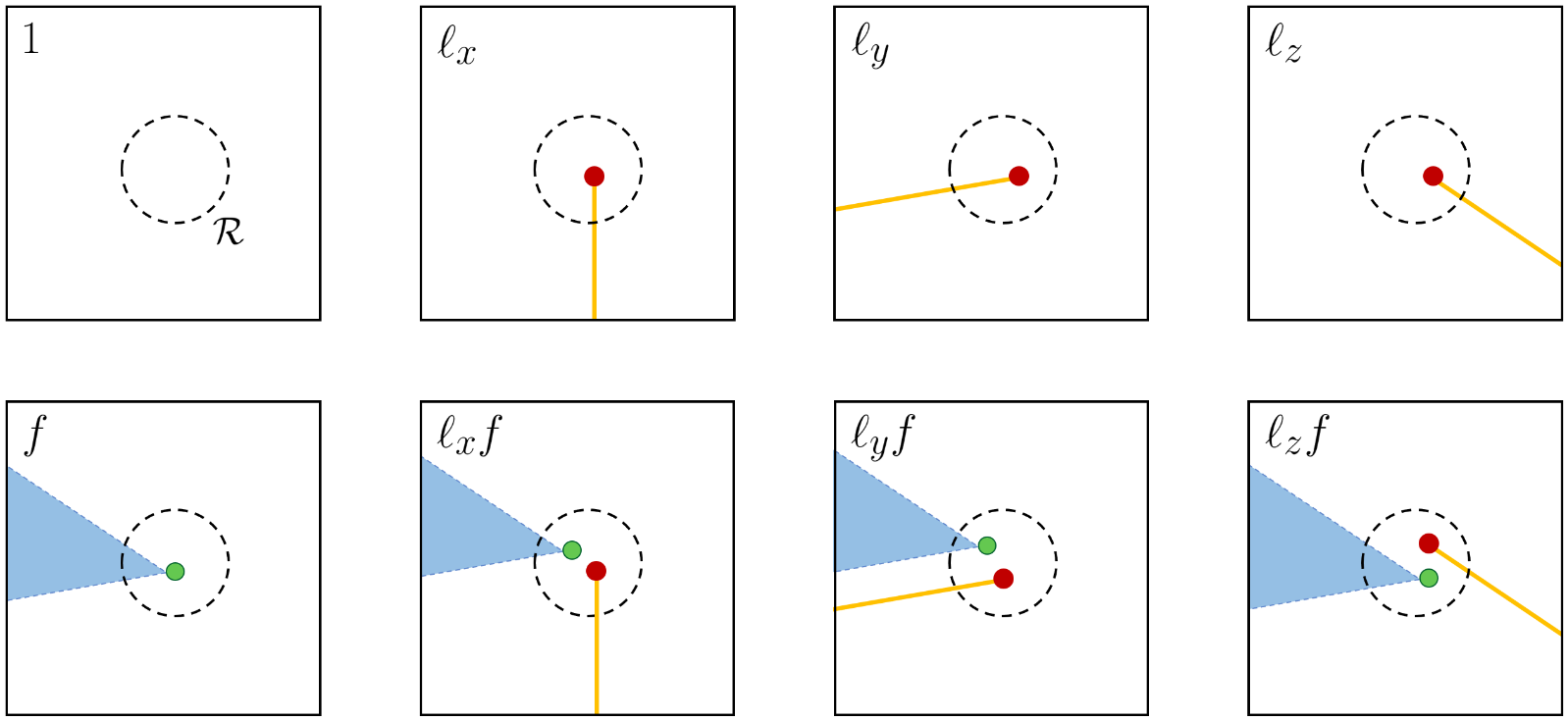}
\caption{The 8 quotient superselection sectors of the X-cube model. The $\ell_x$, $\ell_y$, and $\ell_z$ lineons (red) are created by string operators (orange) along the $x$, $y$, and $z$ direction, respectively. The fracton (green) is created by a membrane operator (blue).}
\label{fig:XCsectors}
\end{figure}

As an example, let us apply this definition to the X-cube model. First consider the lineon excitations of the model, which can move within a straight line via the action of rigid string segment (unitary) operators. The key observation is that lineon dipoles, where the dipolar axis is normal to the direction of mobility, are themselves fractional planon excitations. Thus, the action of the flexible string operators that create these dipolar planons can effectively translate lineon excitations parallel to their dipolar axes (see \figref{fig:XCquotient} for an example). However, there are no such operators capable of transmuting say, an $x$ direction lineon into a $y$ direction lineon. Similarly, the planon creation operators that create fracton dipoles (of any axial orientation $x$, $y$, or $z$), i.e. half-open ribbon operators, are effective hopping operators for individual fractons. Moreover, since these operators nucleate fracton dipoles out of the vacuum, only the total fracton parity of a given state is relevant in determining the quotient superselection sector to which that state belongs. Likewise, only the parities of the number of $x$, $y$, and $z$ direction lineons come into play. However, due to the triple lineon fusion rule, one of these parities is actually redundant; hence, there are a total of 8 quotient sectors for the X-cube model. Let us label the quotient superselection sector with odd fracton parity and even lineon parities as $f$, and the sector with odd parity of direction-$\sigma$ lineons ($\sigma=x,y,z$) and all other parities even as $\ell_\sigma$. Note that due to the triple fusion rule, the sector $\ell_z$ corresponds also to odd parity of $x$ and $y$ direction lineons and even parity of fractons and $z$ direction lineons. Finally, sectors with odd fracton parity, odd parity of $\sigma$ direction lineons, and even parity of the other two types of lineons will be labelled $\ell_\sigma f$. The 8 quotient sectors in $\mathcal{A}$ are thus the vacuum sector 1, the fracton sector $f$, the lineon sectors $\ell_x$, $\ell_y$, and $\ell_z$, and the composite sectors $\ell_x f$, $\ell_y f$, $\ell_z f$. In \figref{fig:XCsectors} we have illustrated representative states of each quotient sector.

The set of quotient superselection sectors for abelian phases has a natural abelian group structure in which the group multiplication corresponds to fusion of quotient sectors, and the vacuum sector represents the identity. To make this precise, we can consider two nearby non-overlapping regions $\mathcal{R}_1$ and $\mathcal{R}_2$ which are encompassed by a larger region $\mathcal{R}$. Suppose $\mathcal{R}_1$ contains excitations of total quotient charge $a$, and $\mathcal{R}_2$ of total quotient charge $b$. Then $a$ and $b$ fuse into $c$, written as $a\times b=c$, if the encompassing region $\mathcal{R}$ has total quotient charge $c$. For the X-cube model, the non-trivial fusion rules are $a\times a=1$ for all $a\in\mathcal{A}$, secondly $\ell_\sigma\times f=\ell_\sigma f$  (where $\sigma=x,y,z$), and finally $\ell_x\times \ell_y\times\ell_z=1$. Hence the fusion group is $\mathbb{Z}_2\times \mathbb{Z}_2\times \mathbb{Z}_2$ with generators $\ell_x$, $\ell_y$, and $f$.

\section{Quasiparticle statistics}

\label{sec:statistics}

In this section, we develop a notion of interferometric detection of quotient superselection sectors, which will serve as the foliated fracton phase analog of quasiparticle braiding statistics in abelian topological phases. The basic idea is to consider equivalence classes of operators which interferometrically detect the presence of excitations in a given quotient sector via Aharanov-Bohm-like phases. Whereas for 2D topological phases, these operators correspond to processes in which one quasiparticle is wound around another, for foliated fracton phases these interferometry operators lack any sort of topological interpretation. Instead, as we will see, they have a geometric character that is inherited from the geometry of the foliation structure.

\subsection{Interferometric detection}

As discussed above, each quotient superselection sector $a\in\mathcal{A}$ corresponds to a subspace $\mathcal{H}_a$ of states containing fractional excitations of quotient charge $a$ within a fixed region $\mathcal{R}$. To formulate a notion of universal quasiparticle statistics for foliated fracton phases, we consider the set $\mathcal{O}$ of \textit{interferometric operators}, which are defined with reference to the region $\mathcal{R}$. An \textit{interferometric operator} is a local unitary operator $O$ that (1) commutes with the Hamiltonian, (2) has compact support in $\overline{\mathcal{R}}$, the complement of $\mathcal{R}$, (3) acts as a pure phase $e^{i\theta_a(O)}$ within each subspace $\mathcal{H}_a$ ($O\ket{a}=e^{i\theta_a(O)}\ket{a}$ for all $\ket{a}\in\mathcal{H}_a$), and  (4) acts as the identity on the ground state ($O\ket{\psi_0}=\ket{\psi_0}$). We note that condition (3) strongly restricts the set of interferometric operators, because states in a given subspace $\mathcal{H}_a$ may differ by the presence of planons in $\mathcal{R}$; hence interferometric operators must be indifferent to these excitations.  Condition (4) merely specifies the overall phase of operators in $\mathcal{O}$.

The set $\mathcal{O}$ may be naturally partitioned into a finite set of classes $\mathcal{O}_i$ ($i\in I$ where $I$ is a finite set), according to equivalence of the statistical phase angles $\theta_a(O)$. That is, if $\theta_a(O)=\theta_{ai}$ for all $a\in\mathcal{A}$, then $O\in\mathcal{O}_i\subset\mathcal{O}$, where the $\theta_{ai}$ have been introduced as statistical angles which depend only on the interferometry class $i$ and the quotient sector $a$. 

These phase factors are the foliated fracton analog of long-range Aharanov-Bohm interactions in abelian 2D topological orders. They arise due to the non-trivial commutation relations between interferometric operators $O$ and operators $W_a$ which create fractional excitations of quotient charge $a$ out of the vacuum: $OW_a=e^{i\theta_{ai}}W_aO$ for all $O\in\mathcal{O}_i$ and $W_a$ such that  $W_a\ket{\psi_0}\in\mathcal{H}_a$. (More precisely, these operators bring excitations of quotient charge $a$ from spatial infinity to the region $\mathcal{R}$.) The statistical phase angles $\theta_{ai}$ are well-defined for generic gapped models, and robust under adiabatic deformation of the Hamiltonian. They are thus universal quantities which partially characterize the foliated fracton phase surrounding a generic model. We note that the set $I$ naturally forms an abelian group with the trivial class as the identity element, and the addition operation coming from operator composition. The maps $i\mapsto e^{i\theta_{ai}}$ are homomorphisms from $I\to U(1)$.

It is interesting to note that in all the models we have considered, the interferometric operators can be thought of as processes in which a dipolar planon of macroscopic dipole length is braided in a 2D plane around the region $\mathcal{R}$. Moreover, we find in all cases that the number of interferometric classes is equal to the number of QSS. Thus, it is natural to arrange the statistical phases in matrix form: $S_{ai}=e^{i\theta_{ai}}$. In fact, the resulting $S$ matrix is a direct generalization of the topological $S$ matrix in the theory of 2D topological orders, in the sense that the equivalent definitions applied to 2D topological phases yields the topological $S$ matrix. However, the $S$ matrix for foliated fracton phases differs from the topological $S$ matrix in that there is no inherent symmetry between the row and column indices, whereas for the topological $S$ matrix both indices correspond to anyon species and it generically holds that $S_{ab}=S^*_{ba}$.

\begin{figure}[htbp]
\centering
\includegraphics[width=.35\textwidth]{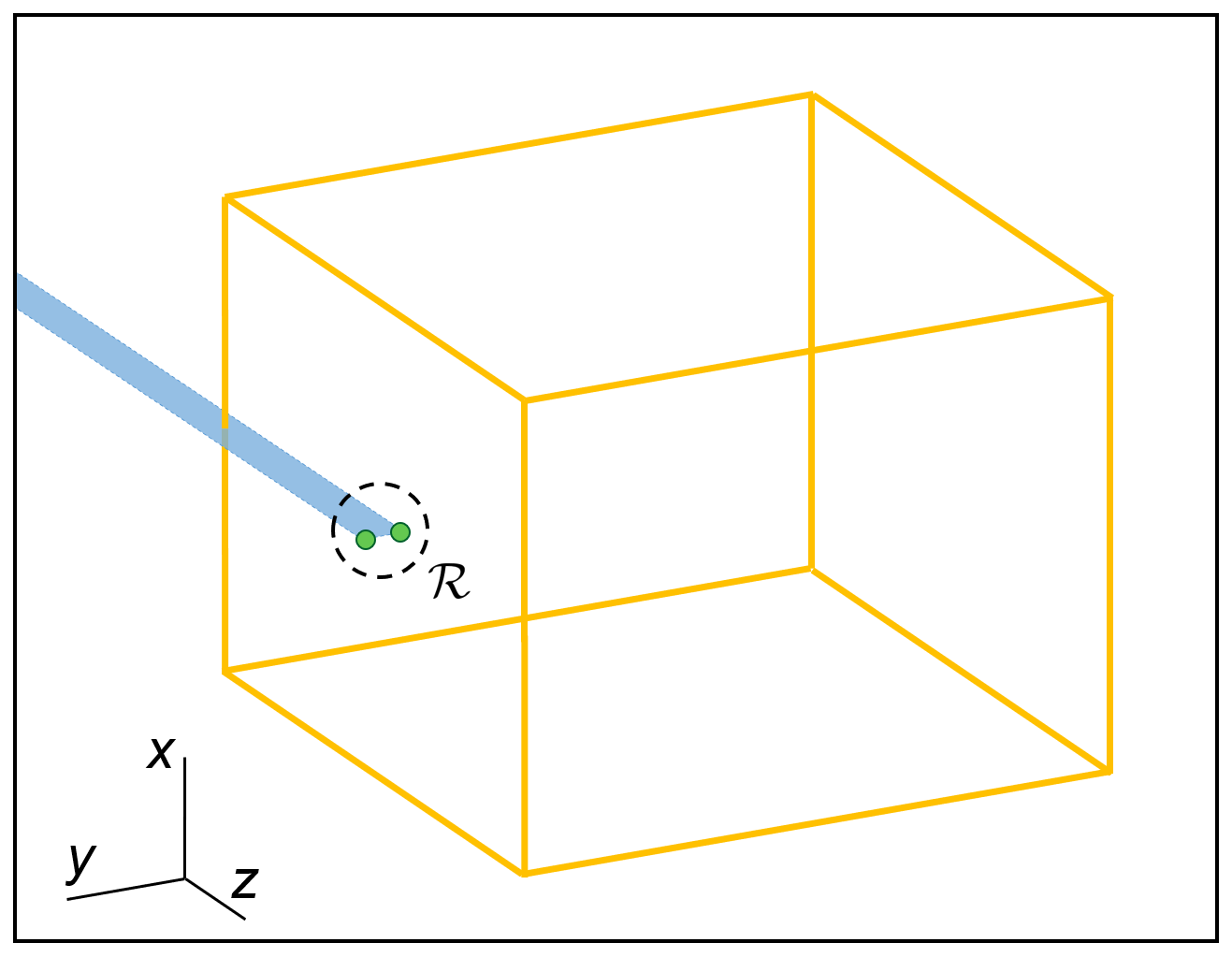}
\caption{An example of a wireframe operator $O$ (defined as the tensor product of Pauli $X$ operators along the yellow edges) which violates condition (3) discussed in the main text. This operator anti-commutes with the blue ribbon operator pictured, which creates a planon (i.e. a fracton dipole) in region $\mathcal{R}$. Hence $O$ acts as $+1$ on some states and $-1$ on other states in the trivial quotient superselection sector, and so $O\notin\mathcal{O}$.}
\label{fig:condition3}
\end{figure}

\subsection{Example: X-cube model}

Let us continue to consider the X-cube model as a primary example. First, we need to determine the set of interferometric operators $\mathcal{O}$ with reference to a particular region $\mathcal{R}$ of an X-cube system. Since the X-cube Hamiltonian is a stabilizer code, the algebra of observables that commute with the Hamiltonian is generated by the Hamiltonian terms themselves. Of the operators that commute with $H$ and have compact support in $\overline{\mathcal{R}}$, condition (3) further restricts this set, because a rigid string operator lying in a plane that intersects $\mathcal{R}$ will anti-commute with certain ribbon operators that create fracton dipoles in $\mathcal{R}$. An example of an operator which satisfies conditions (1) and (2) but not (3) is shown in \figref{fig:condition3}. It can then be seen by careful inspection that the set $\mathcal{O}$ contains 8 inequivalent classes of interferometric operators (including the trivial class).

\begin{figure}[htbp]
\centering
\includegraphics[width=.95\textwidth]{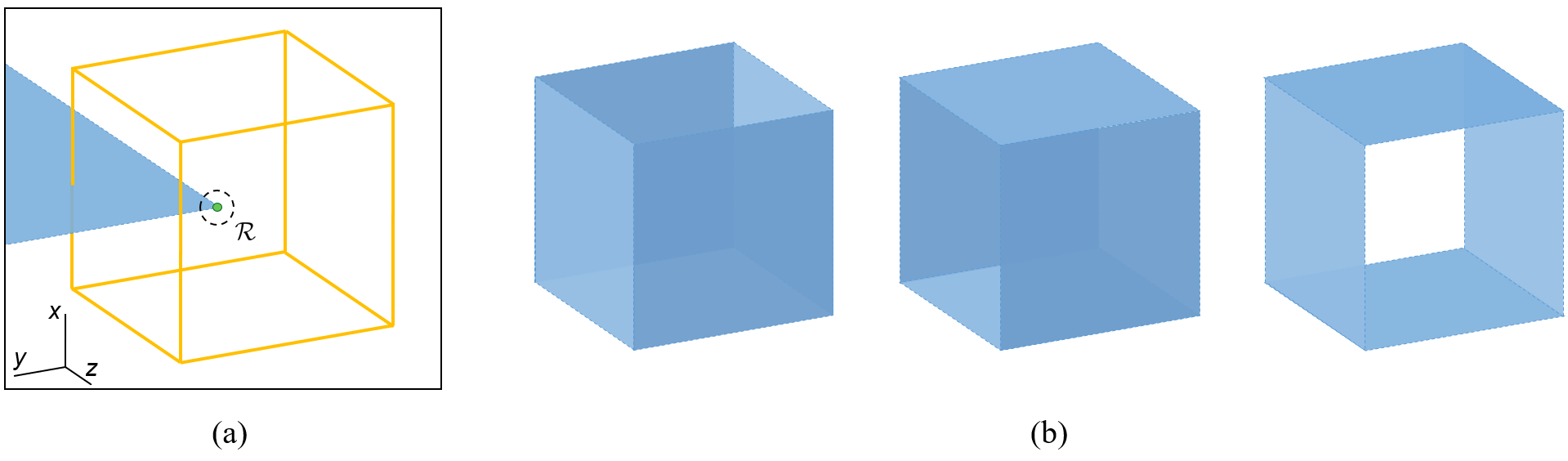}
\caption{(a) An example of a wireframe interferometric operator belonging to the $F$ class, defined as the product of Pauli $X$ operators over the yellow edges. The region $\mathcal{R}$ is located at the center of the wireframe. This operator detects the presence of a fracton (green dot) due to the anti-commutation relation with the blue membrane operator. (b) Cylindrical membrane operators representing, from left to right, the $X$, $Y$ and $Z$ interferometry classes, of which only 2 are independent. In each case $\mathcal{R}$ resides at the center of the prism.}
\label{fig:XCbraiding}
\end{figure}

The first non-trivial class, denoted by the label $F$, contains wireframe string operators that measure the fracton parity in region $\mathcal{R}$, and are insensitive to the presence of lineons. An example is shown in \sfigref{fig:XCbraiding}{a}. In other words, $\theta_{f,F}=\theta_{\ell_xf,F}=\theta_{\ell_yf,F}=\theta_{\ell_zf,F}=\pi$, whereas the remaining phase factors are trivial. The $e^{i\pi}$ phase factors arise due to the anti-commutation relation between these wireframe operators and the membrane operators that create fractons in $\mathcal{R}$ at their corners. Conversely, the next three non-trivial classes, denoted by the labels $X$, $Y$, and $Z$, detect lineon parity and are insensitive to the fracton sector. They obey the relation $\mathcal{O}_X\mathcal{O}_Y\mathcal{O}_Z=\mathcal{O}_1$. The $X$ class contains large membrane operators with cylindrical topology which wrap around $\mathcal{R}$ around the $x$ axis, and likewise for the $Y$ and $Z$ classes. These membrane operators anti-commute with the half-open rigid string operators that create lineons of quotient charge $\ell_y$ and $\ell_z$ in $\mathcal{R}$, while commuting with those that create lineons of charge $\ell_x$. Therefore, $\theta_{\ell_y,X}=\theta_{\ell_z,X}=\pi$ whereas $\theta_{\ell_x,X}=1$, and similarly for the fracton-lineon composite sectors (and likewise for cyclic permutations of the indices). Examples of these cylindrical operators are shown in \sfigref{fig:XCbraiding}{b}. It is instructive to note that the four interferometric classes indexed by $X$, $Y$, $Z$, and $F$, respectively detect violations of the $A^{yz}_v$, $A^{xz}_v$, $A^{xy}_v$, and $B_c$ terms of the X-cube Hamiltonian in \eqnref{eqn:XcubeH} (or rather, odd numbers of violations). The remaining three classes, given the labels $XF$, $YF$, and $ZF$, contain compositions of operators in the other classes. The group structure on the classes of interferometric operators for the X-cube model is therefore $\mathbb{Z}_2\times \mathbb{Z}_2\times \mathbb{Z}_2$ with generators $X$, $Y$, and $F$. The statistics are summarized in the following $S$ matrix, in which the rows correspond to QSS in the order $\{1, \ell_x, \ell_y, \ell_z, f, \ell_xf, \ell_yf, \ell_zf\}$, and the columns correspond to interferometric classes in the order $\{1, X, Y, Z, F, XF, YF, ZF\}$:
\begin{equation}
    S=
    \begin{pmatrix}
        1 & 1  &   1 & 1  & 1  & 1  & 1  & 1 \\
        1 & 1  &  -1 & -1 & 1  & 1  & -1 & -1 \\
        1 & -1 & 1   & -1 & 1  & -1 & 1  & -1 \\
        1 & -1 & -1  & 1  & 1  & -1 & -1 & 1 \\
        1 & 1  & 1   & 1  & -1 & -1 & -1 & -1 \\
        1 & 1  & -1  & -1 & -1 & -1 & 1  & 1 \\
        1 & -1 & 1   & -1 & -1 & 1  & -1 & 1 \\
        1 & -1 & -1  & 1  & -1 & 1  & 1  & -1 \\
    \end{pmatrix}.
    \label{SXc}
\end{equation}

\section{More examples}

\label{sec:examples}

In this section we examine the quotient superselection sectors, fusion rules, and classes of interferometric operators and statistical phases for a handful of exactly solvable models, which may be viewed as RG fixed point representatives of corresponding foliated fracton phases. In \secref{sec:anisotropic} we introduce a novel anisotropic foliated fracton model which exhibits fractional lineon and planon excitations but no fractons.

\subsection{Stack of 2D topological orders}

A simple decoupled stack of 2D topological orders, viewed as a 3D model, belongs to the trivial foliated fracton phase according to the definition proposed in Ref. \cite{3manifolds}. This is reflected in the structure of excitations in such models: since all fractional excitations are anyons of the constituent layers, there is only one quotient superselection sector, that of the vacuum.

\subsection{Semionic X-cube model}
\label{sec:SXc}

\begin{figure}[htbp]
\centering
\includegraphics[width=0.95\textwidth]{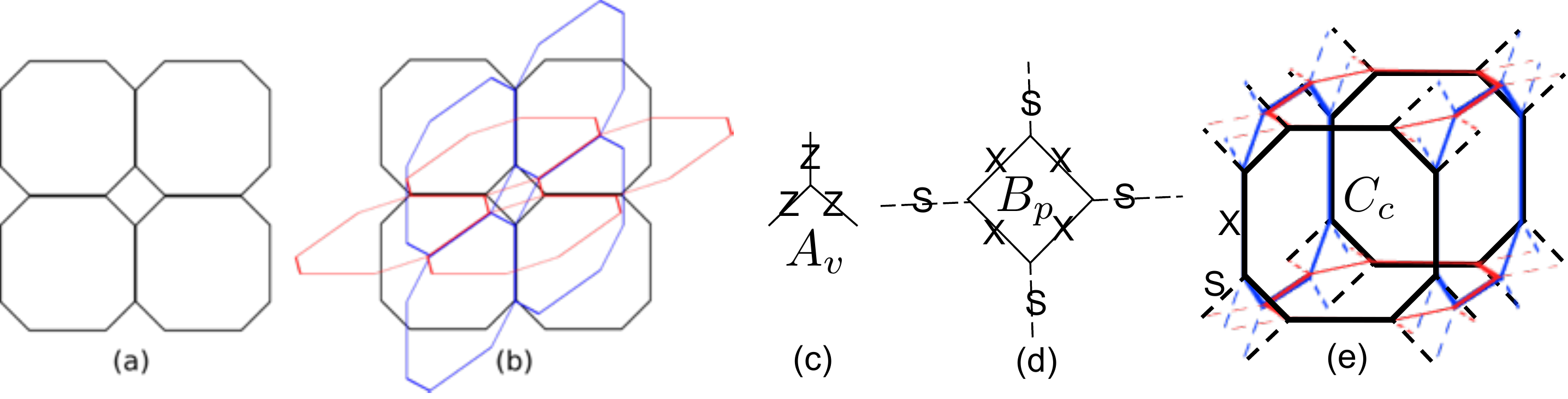}
\caption{(a) A trivalent 2D lattice obtained by decorating a 2D square lattice with diamond plaquettes at each vertex. (b) Stacks of such trivalent lattices in the $xy$, $yz$, and $zx$ planes; the edges in $x$, $y$ and $z$ directions overlap in pairs. (c) The vertex term, (d) the diamond plaquette term, and (e) the cube term of the semionic X-cube Hamiltonian. Here, $X$ and $Z$ are Pauli operators and $S = \text{diag}(1,i)$. In the cube term (e), there is one $X$ on each qubit on the solid edges and one $S$ on each qubit on the dashed edges. For clarity, we only draw one of each.}
\label{fig:triv_stack}
\end{figure}

The semionic X-cube model was discussed in Ref.~\cite{MaLayers} as a semionic generalization of the original X-cube model. The model is defined on a variation of the cubic lattice which can be obtained as the union of three stacks of 2D decorated square lattices parallel to the $xy$, $yz$ and $zx$ planes (\sfigref{fig:triv_stack}{b}). In each 2D plane, a small diamond shape is added at each vertex of the square lattice so that in the new lattice each vertex has degree three (\sfigref{fig:triv_stack}{a}). The Hamiltonian contains three types of terms: a vertex term $A_v$ at each of the trivalent vertices in the $xy$, $yz$ and $zx$ planes as shown in \sfigref{fig:triv_stack}{c}, a plaquette term $B_p$ at each diamond plaquette in the planes as shown in \sfigref{fig:triv_stack}{d}, and a cube term $C_c$ at each cubic cell as shown in \sfigref{fig:triv_stack}{e}: 
\begin{equation}
H_1 = -\sum_v A^{(1)}_v - \sum_p B^{(1)}_p - \sum_c C^{(1)}_c.
\label{HsXC}
\end{equation}

For comparison, we can also define the X-cube model on the decorated lattice:
\begin{equation}
H_0 = -\sum_v A^{(0)}_v - \sum_p B^{(0)}_p - \sum_c C^{(0)}_c
\label{HXC}
\end{equation}
The Hamiltonian also takes the form shown in \sfigref{fig:triv_stack}{c-e} but differs from $H_1$ in that the operator $S$ is absent from the dashed lines.
As explained in Ref. ~\cite{MaLayers}, the X-cube and semionic X-cube model on the decorated lattice can be obtained by taking toric code or double semion layers respectively in each $xy$, $yz$ and $zx$ plane and coupling them together. The $A_v$ and $B_p$ terms come directly from the vertex and plaquette terms of the toric code and double semion models. The $C_c$ term is a combination of six plaquette terms on neighboring planes. 

The quotient superselection sectors and the $S$ matrix of the X-cube model on the decorated lattice is the same as those on the original cubic lattice. To see this, we note that on the decorated lattice, violations of the $A_v$ and $C_c$ terms correspond to the lineon and fracton excitations as before while violations of the $B_p$ term are a new type of planon. Interferometric operators take the same form as before (wireframe, cylinder and their composition) except for the decoration at each vertex. There are still eight quotient superselection sectors and eight interferometric operators which give rise to the same $S$ matrix as in Eq.~\ref{SXc}.

Similarly, the semionic X-cube model on the decorated cubic lattice has eight quotient superselection sectors generated by a fracton and two lineons as discussed in Ref.~\cite{MaLayers}. To detect these sectors interferometrically, we can use a wireframe shaped operator which is a composition of all $C_c$ operators inside the wireframe. There are also cylinder shaped interferometric operators and they take the same form as in the X-cube model. Direct calculation shows that the $S$ matrix of the semionic X-cube model is the same as that of the X-cube model (Eq.~\ref{SXc}).

\subsection{Stacked kagome lattice X-cube model}
\label{sec:KXc}

As discussed in Ref. \cite{Slagle17Lattices}, it is possible to define the X-cube model on generalized lattices which arise as the triple intersection points of three or more stacks of parallel planes. This class includes the stacked kagome lattice, which is formed from 4 underlying stacks. These stacks are normal to the $(0,1,0)$, $\left(\sqrt{3}/2,1/2,0\right)$, $\left(-\sqrt{3}/2,1/2,0\right)$, and $(0,0,1)$ directions respectively. The fourth stack, whose layers are parallel to the $xy$ plane, contains embedded 2D kagome lattices. Actually, these stacks represent an underlying foliation structure whose leaves correspond to 2D toric code layers; hence the stacked kagome X-cube model constitutes a foliated fracton model composed of 4 foliations. As in the normal X-cube model, qubits are placed on each edge of the lattice, and the Hamiltonian takes the form
\begin{equation}
    H=-\sum_{v}(A^1_v+A^2_v+A^3_v)-\sum_{c}B_c.
    \label{eqn:kagomeXCH}
\end{equation}
In this case, $v$ runs over all vertices, and the operators $A^i_v$ are tensor products of Pauli $Z$ operators over four coplanar edges adjacent to $v$, one for each of the 3 foliating planes containing $v$. Here, $c$ indexes the elementary 3-cells of the lattice, which are all either triangular or hexagonal prisms, and $B_c$ is a tensor product of Pauli $X$ operators over the edges of $c$.

\begin{figure}[htbp]
\centering
\includegraphics[width=0.4\textwidth]{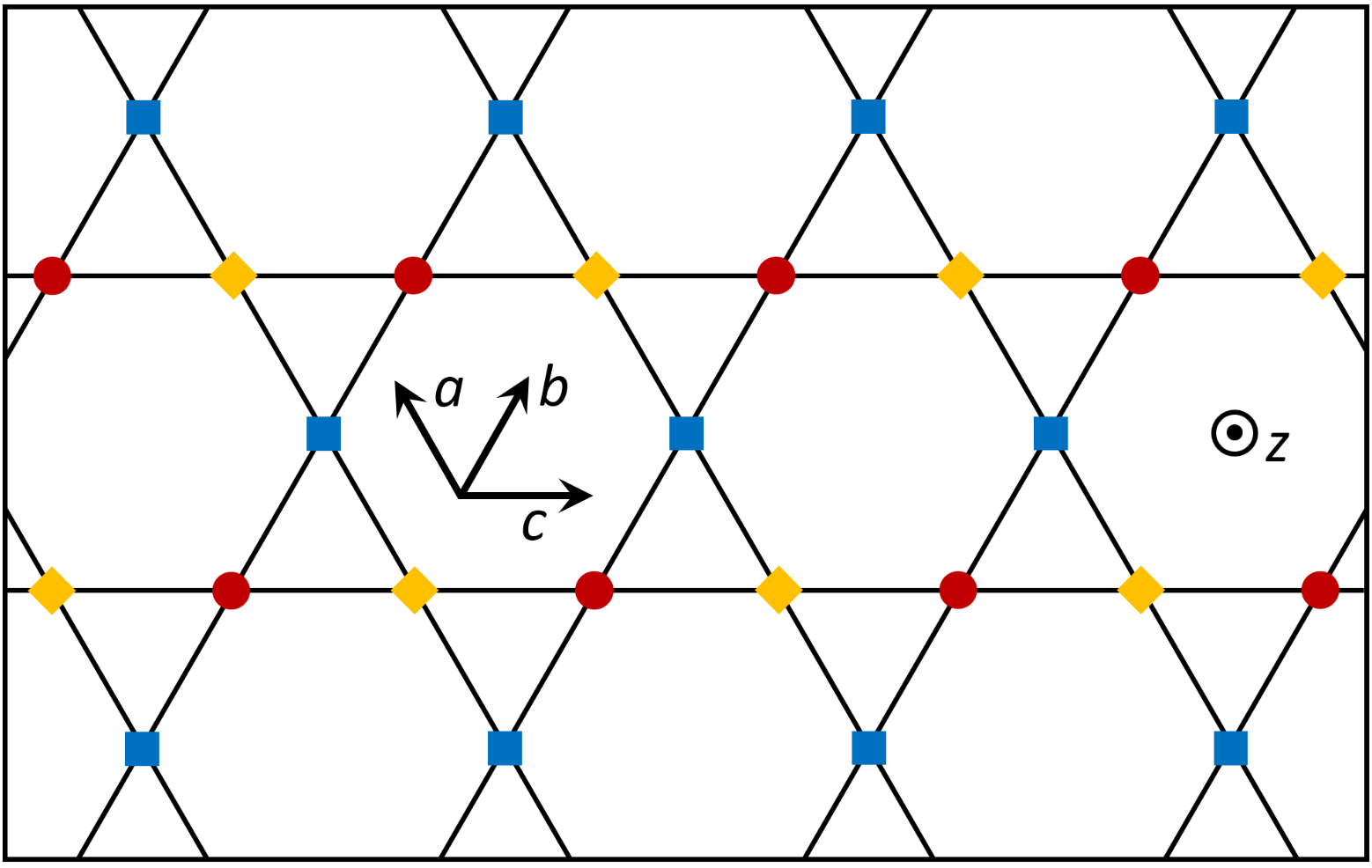}
\caption{The three sublattices of the kagome lattice, indicated by red circles, blue squares, and yellow diamonds. In the stacked kagome X-cube model, there are three types of $z$-direction (out of the plane) lineons, corresponding to the three sublattices.}
\label{fig:kagomeLineons}
\end{figure}

The excitation structure of the stacked kagome X-cube model is quite similar in spirit to that of the original X-cube model: violations of the vertex Hamiltonian terms are lineons, whereas violations of the 3-cell terms are fractons. However, there are four possible directions of mobility for lineons: $a=\left(-1/2,\sqrt{3}/2,0\right)$, $b=\left(1/2,\sqrt{3}/2,0\right)$, $c=(1,0,0)$ and $\hat{z}=(0,0,1)$. (The first 3 directions lie within the 2D kagome layers, whereas the fourth is normal to them). In all cases, the lineons are mobile along the line of intersection of two of the underlying foliation leaves. Pairs of $a$, $b$, or $c$ direction lineons separated along the $z$ axis or within the $xy$ plane constitute fractional planon excitations. Thus the lineons mobile in each of these 3 directions constitute their own quotient superselection sectors, which we will label $\ell_a$, $\ell_b$, and $\ell_c$. On the other hand, the $z$ direction lineons can be divided into three sublattices as shown in \figref{fig:kagomeLineons}; lineon dipoles may be free to move in a 2D plane only if the two lineons belong to the same sublattice. Thus, each of these three types of $z$ direction lineons represents a quotient superselection sector as well, labelled $\ell_R$, $\ell_Y$, and $\ell_B$. However, due to the triple fusion rules, each of these quotient sectors is the result of fusion of two of the sectors $\ell_a$, $\ell_b$, and $\ell_c$. In particular, the fusion rules are
\begin{align*}
    \ell_R&=\ell_b\times\ell_c \\
    \ell_Y&=\ell_a\times\ell_c \\
    \ell_B&=\ell_a\times\ell_b.
\end{align*}
Therefore, there are only 3 independent lineon quotient sectors. The last non-trivial lineon QSS is given by the fusion result $\ell_a\times\ell_b\times\ell_c$.

\begin{figure}[htbp]
\centering
\includegraphics[width=0.95\textwidth]{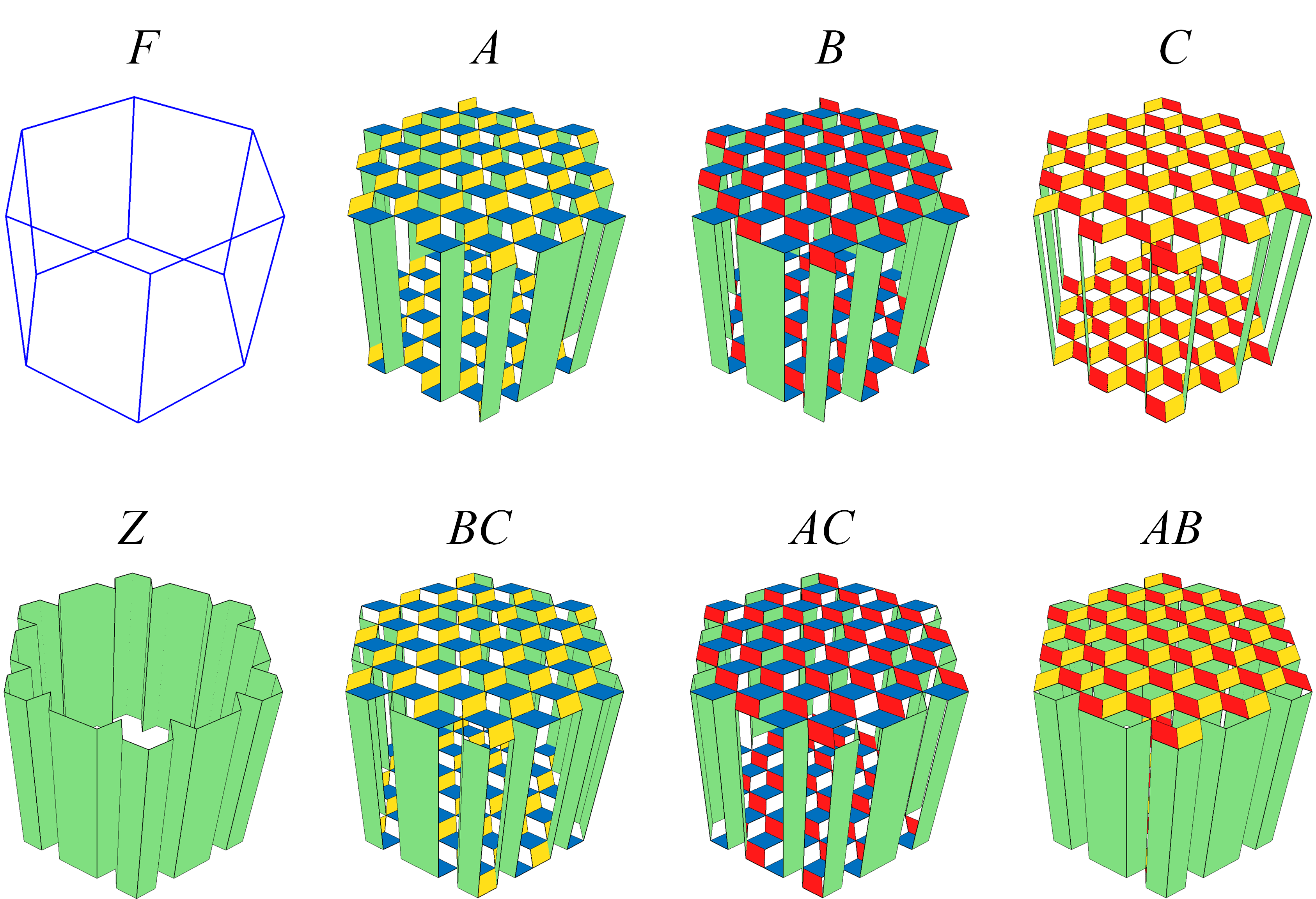}
\caption{Representative operators of the interferometric classes in the stacked kagome X-cube model. The top left figure depicts a wireframe operator which is a tensor product of Pauli $X$ operators over the qubits along the edges. The remaining figures depict membrane operators which are tensor products of Pauli $Z$ operators over the dual lattice plaquettes drawn in the figures. These operators may also be chosen to be in the shape of a triangular prism.}
\label{fig:SKXCoperators}
\end{figure}

As in the cubic lattice X-cube model, dipoles of adjacent fractons are themselves planons, and thus all fractons belong to the same quotient superselection sector. In total there are therefore $2^4=16$ quotient superselection sectors in the stacked kagome X-cube model. The group of interferometric operators that detect these sectors are generated by a class $F$ of wireframe operators which detects fractons, as well as 3 independent classes of membrane operators $A$, $B$, and $C$ which detect the presence of lineons. These operators are membrane-like in the sense that they have support along the surface of a polyhedron, which can be chosen to be a hexagonal (or triangular) prism (see \figref{fig:SKXCoperators}). The remaining classes of membrane operators are $AB=A\times B$, $BC=B\times C$, $AC=A\times C$, and $Z=A\times B\times C$.

Since the fracton and lineon sectors are independent of each other in this model, it is instructive to construct an abbreviated $\tilde{S}$ matrix which contains the interferometric statistics between the lineon QSS and membrane operators alone. Indexing the rows in the order $\{1,\ell_a,\ell_b,\ell_c,\ell_a\times\ell_b\times\ell_c,\ell_R,\ell_Y,\ell_B\}$ and the columns in the order $\{1,A,B,C,Z,BC,AC,AB\}$, this matrix takes the form
\begin{equation}
    \tilde{S}=
    \begin{pmatrix}
        1 & 1  &   1 & 1  & 1  & 1  & 1  & 1 \\
        1 & -1 &   1 & 1  &-1  & 1  & -1 & -1 \\
        1 &  1 & -1  &  1 & -1 & -1 & 1  & -1 \\
        1 &  1 &  1  & -1 & -1 & -1 & -1 & 1 \\
        1 & -1 & -1  & -1 & -1 &  1 &  1 & 1 \\
        1 & 1  & -1  & -1 &  1 &  1 & -1 & -1 \\
        1 & -1 & 1   & -1 &  1 & -1 &  1 & -1 \\
        1 & -1 & -1  & 1  &  1 & -1 & -1 & 1 \\
    \end{pmatrix}.
\end{equation}
The full $S$ matrix including the fracton QSS and wireframe interferometric operators then takes the form
\begin{equation}
    S=\begin{pmatrix}
        1 & 1 \\
        1 & -1 \\
    \end{pmatrix}\otimes\tilde{S}.
\end{equation}

\subsection{Hyperkagome lattice X-cube model}
\label{sec:HKXc}

\begin{figure}[htbp]
\centering
\includegraphics[width=0.5\textwidth]{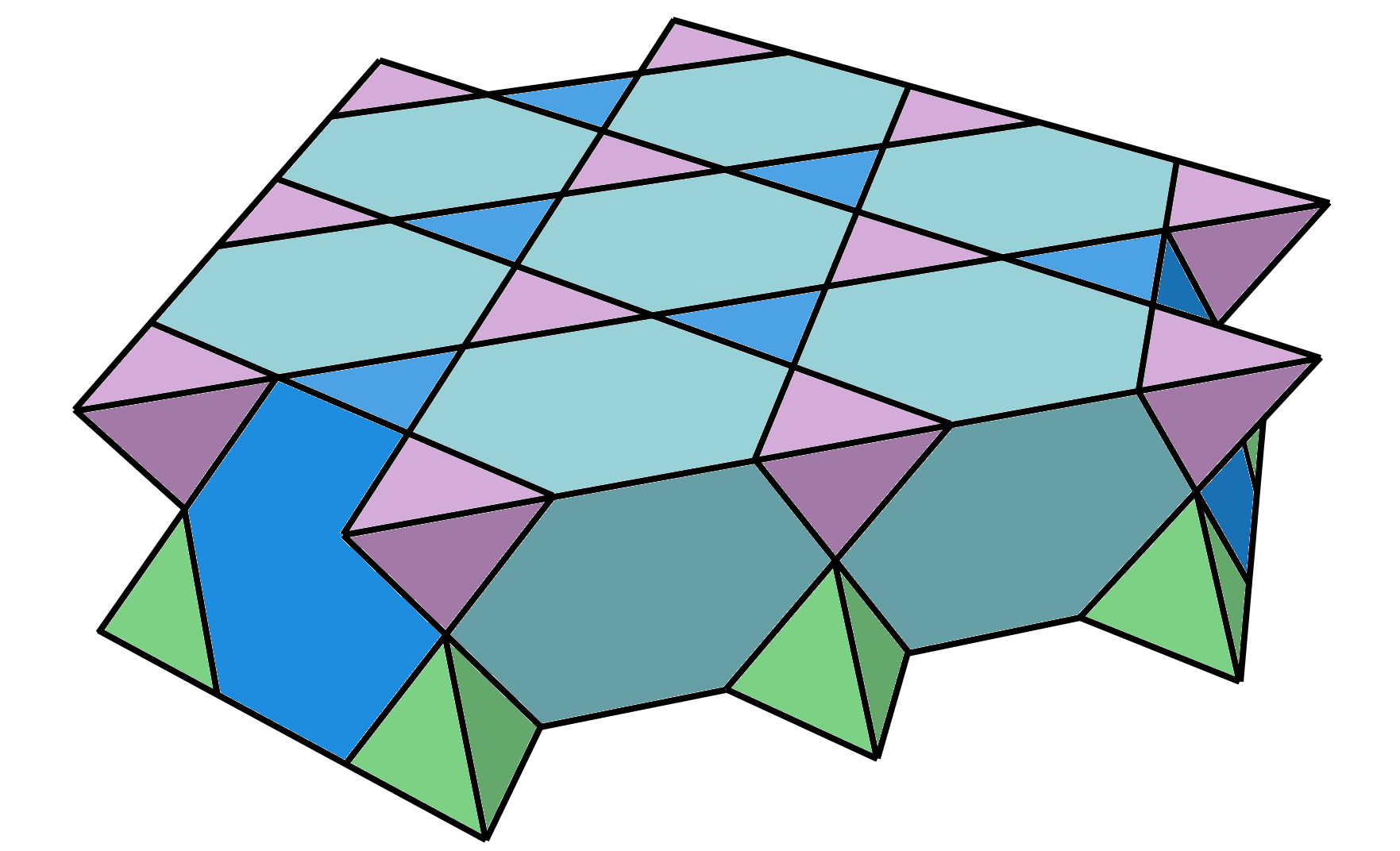}
\caption{The hyperkagome lattice. The elementary 3-cells are small  tetrahedra (green and purple) and large truncated tetrahedra (turqoise and blue).}
\label{fig:hyperkagome}
\end{figure}

Using an analogous construction to the previous example, it is possible to define a version of the X-cube model where the qubits reside on the edges of a hyperkagome lattice (also known as a quarter cubic honeycomb lattice). The hyperkagome lattice arises as the set of triple intersection points of planes belonging to the 4 discrete foliations defined by the equations $-x+y+z=k$, $+x-y+z=k$, $+x+y-z=k$, and $+x+y+z=k+1/2$ for all $k\in\mathbb{Z}$. The Hamiltonian for this version of the X-cube model takes the same form as in \eqnref{eqn:kagomeXCH}; for the hyperkagome lattice, the elementary 3-cells consist of small 
tetrahedra and large truncated tetrahedra, as shown in \figref{fig:hyperkagome}.

\begin{figure}[htbp]
\centering
\includegraphics[width=0.85\textwidth]{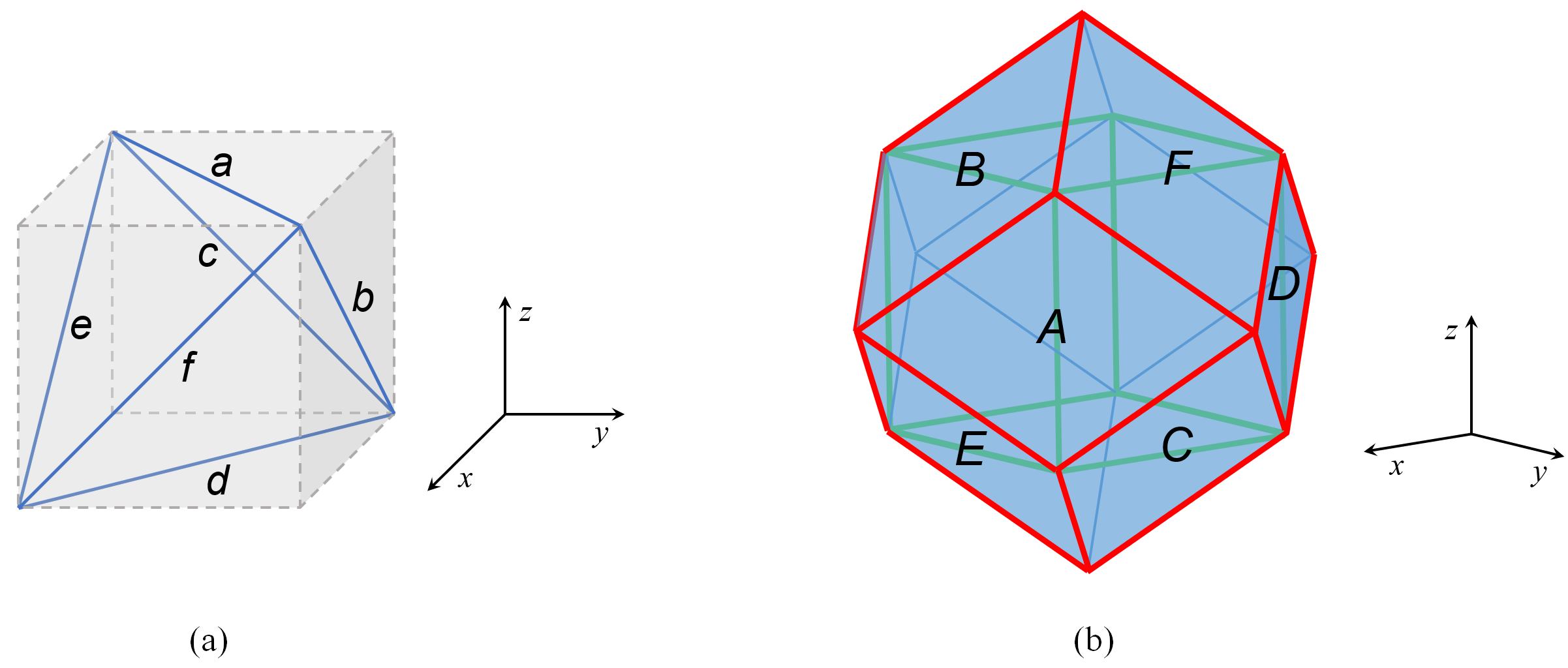}
\caption{(a) The six possible directions of mobility of lineons in the hyperkagome X-cube model. (b) A rhombic dodecahedron. Each face of the dodecahedron is normal to one of the 6 directions of lineon mobility.}
\label{fig:polyhedra}
\end{figure}

In this geometry, there are 6 species of lineons which move along the $a=(1,1,0)$, $b=(1,0,1)$, $c=(0,1,-1)$, $d=(1,-1,0)$, $e=(1,0,-1)$, and $f=(0,1,1)$ directions, corresponding to lines of intersection of the foliating planes (see \sfigref{fig:polyhedra}{a}). As in the other X-cube models, pairs of lineons moving in the same direction may combine to form dipolar planon excitations. Thus, all lineons mobile in the $\sigma$ direction belong to a single quotient superselection sector $\ell_\sigma$. Moreover, there are four triple fusion rules:
\begin{align*}
    \ell_a\times \ell_b\times\ell_f&=1 \\
    \ell_b\times\ell_c\times\ell_d&=1 \\
    \ell_a\times\ell_c\times\ell_e&=1 \\
    \ell_d\times\ell_e\times\ell_f&=1.
\end{align*}
Therefore, there are exactly 3 independent lineon quotient sectors, which can be chosen to be, for instance, $a$, $b$, and $c$. The fusion result $\ell_a\times\ell_b\times\ell_c$ constitutes a 7th non-trivial lineon sector. On the other hand, there is just a single fracton sector $f$ since neighboring fractons combine to form planons as in the cubic and stacked kagome lattice X-cube models. Hence there are a total of $2^4=16$ QSS.

As in the stacked kagome X-cube model, in the hyperkagome X-cube model there is one class of wireframe interferometric operators which detects fracton parity, and 3 classes of independent membrane operators which are sensitive to the lineon content of the region $\mathcal{R}$. The membrane operators can be chosen to have support over the surface of a rhombic dodecahedron which is aligned with the Wigner-Seitz cell of the underlying fcc Bravais lattice (itself a rhombic dodecahedron). They can be constructed in the following way. First, note that the cross-shaped Hamiltonian terms correspond to intersections of pairs of lines of lineon mobility, and may be divided into 12 groups and labelled according to the directions of these two lines. For example, if vertex $v$ lies at the intersection of lines oriented in the $a$, $c$, and $e$ directions, then the vertex terms associated with $v$ are $A_v^{ac}$, $A_v^{ce}$, and $A_v^{ae}$.

The membrane operators are then constructed as a product of vertex terms within a large rhombic dodecahedral region $\mathcal{D}$. The microscopic region $\mathcal{R}$ lies at the center of this dodecahedron. In particular, we define
\begin{equation}
    O_{ABC}=\prod_{v\in\mathcal{D}} A_v^{ab}A_v^{bc}A_v^{ac}
\end{equation}
and likewise for $O_{AEF}$, $O_{BDF}$, and $O_{CDE}$. Moreover,
\begin{align*}
    O_{BCEF}&=O_{ABC}O_{AEF} \\
    O_{ACDF}&=O_{ABC}O_{BDF} \\
    O_{ABDE}&=O_{ABC}O_{CDE}.
\end{align*}
A rigid string operator that creates a lineon in region $\mathcal{R}$ must pierce the center of one of the 12 faces of $\mathcal{D}$ (see \figref{fig:polyhedra}{b}). The interferometric operators are constructed such that they anti-commute with rigid string operators passing through some, but not all of these faces. For instance, the operator $O_{ABC}$ anti-commutes with rigid string operators oriented in the $a$, $b$, and $c$ directions. Thus, the abbreviated $\tilde{S}$ matrix, which contains the statistics of the lineon QSS and membrane interferometry operators, with respect to the bases $\{1,\ell_a,\ell_b,\ell_c,\ell_a\times\ell_b\times\ell_c,\ell_d,\ell_e,\ell_f\}$ and $\{1,AEF,BDF,CDE,ABC,BCEF,ACDF,ABDE\}$, takes the form
\begin{equation}
    \tilde{S}=
    \begin{pmatrix}
        1 & 1  &   1 & 1  & 1  & 1  & 1  & 1 \\
        1 & -1 &   1 & 1  &-1  & 1  & -1 & -1 \\
        1 &  1 & -1  &  1 & -1 & -1 & 1  & -1 \\
        1 &  1 &  1  & -1 & -1 & -1 & -1 & 1 \\
        1 & -1 & -1  & -1 & -1 &  1 &  1 & 1 \\
        1 & 1  & -1  & -1 &  1 &  1 & -1 & -1 \\
        1 & -1 & 1   & -1 &  1 & -1 &  1 & -1 \\
        1 & -1 & -1  & 1  &  1 & -1 & -1 & 1 \\
    \end{pmatrix}.
\end{equation}
and the full $S$ matrix is given by
\begin{equation}
    S=\begin{pmatrix}
        1 & 1 \\
        1 & -1 \\
    \end{pmatrix}\otimes\tilde{S}.
\end{equation}
Interestingly, the QSS and quasiparticle statistics of the hyperkagome and stacked kagome X-cube models have identical algebraic structure; the models differ only in the geometry of their foliation structures.

\subsection{$Z_N$ X-cube model}
\label{sec:ZNXc}

\begin{figure}[htbp]
\centering
\includegraphics[width=0.75\textwidth]{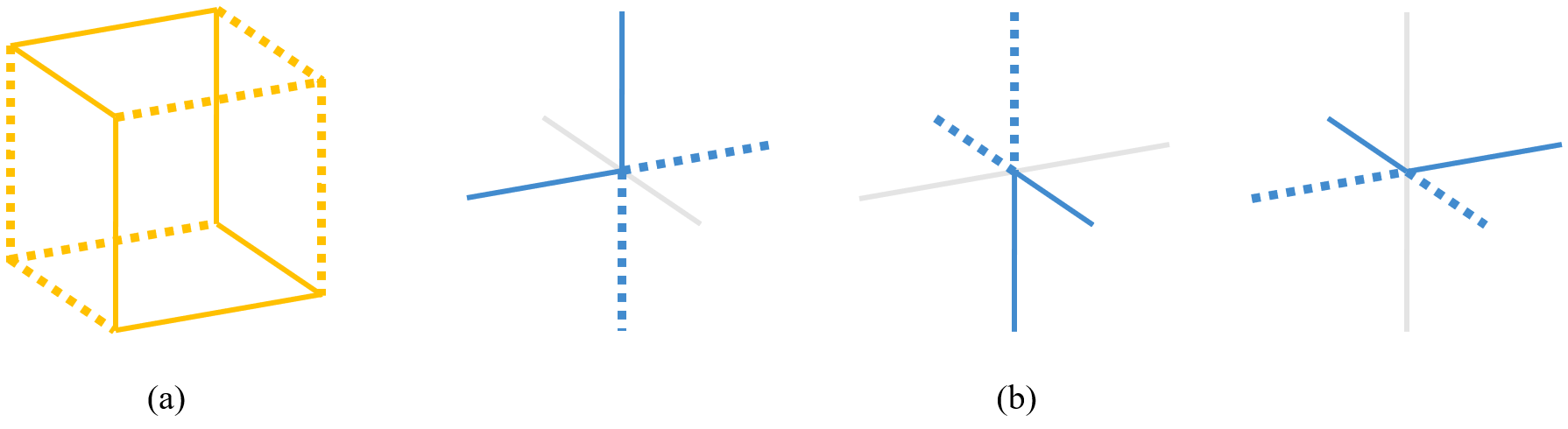}
\caption{(a) Cube term $B_c$ of the $\mathbb{Z}_N$ X-cube Hamiltonian, defined as the tensor product of generalized $Z$ operators over the solid yellow edges and $Z^\dagger$ operators over the dotted yellow edges. (b) Cross stabilizers $A_v^{xy}$, $A_v^{xz}$, and $A_v^{yz}$. They act as generalized $X$ on the solid blue edges and $X^\dagger$ on the dotted blue edges.}
\label{fig:ZNH}
\end{figure}

The X-cube model is also readily generalized to a family of abelian rotor models, in which each edge of a cubic lattice contains a $\mathbb{Z}_N$ rotor degree of freedom spanned by basis states $\ket{0},\ldots,\ket{N-1}$. The Hamiltonian is defined in terms of clock and shift operators $X$ and $Z$ which act as $Z\ket{m}=\omega^m\ket{m}$ and $X\ket{m}=\ket{m+1\hspace{.1cm}\mathrm{mod}\hspace{.1cm}N}$, where $\omega=e^{2\pi i/N}$, and satisfy the commutation relations $ZX=\omega XZ$ and $Z^\dagger X=\omega^{-1} XZ^\dagger$. The Hamiltonian takes the form
\begin{equation}
    H=-\sum_v\left(A^{xy}_v+A^{yz}_v+A^{xz}_v+\mathrm{h.c.}\right)
        -\sum_c\left(B_c+B_c^\dagger\right)
\end{equation}
where per usual $v$ and $c$ run over the vertices and elementary cubes of the lattice, respectively. The operators $A^{xy}_v$, $A^{yz}_v$, $A^{xz}_v$, and $B_c$ are depicted in \figref{fig:ZNH}.

\begin{figure}[htbp]
\centering
\includegraphics[width=0.35\textwidth]{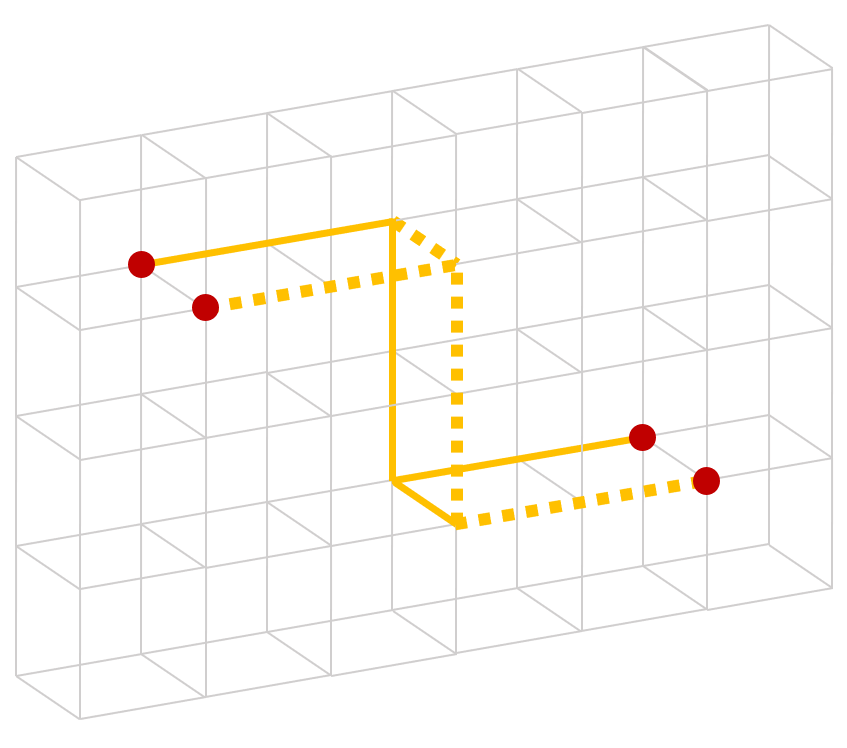}
\caption{A flexible string operator for the $\mathbb{Z}_N$ X-cube model. It is defined as the tensor product of generalized $Z$ operators over the solid yellow edges and $Z^\dagger$ operators over the dotted yellow edges, and creates pairs of lineons, represented as red dots, at its endpoints.}
\label{fig:ZNplanon}
\end{figure}

Like the original $\mathbb{Z}_2$ version, the $\mathbb{Z}_N$ X-cube Hamiltonian is exactly solvable, and exhibits lineon excitations created at the endpoints of rigid string operators, and fracton excitations created at the corners of membrane operators. However, in the rotor model these excitations obey $\mathbb{Z}_N$ fusion rules, and the lineons obey generalized triple fusion rules. Moreover, pairs of adjacent fractons form composite dipolar planons free to move in a 2D plane, as do pairs of adjacent lineons (for example, see \figref{fig:ZNplanon}). As a result, the quotient superselection sectors for the $\mathbb{Z}_N$ X-cube model represent the group $\mathbb{Z}_N\times \mathbb{Z}_N\times \mathbb{Z}_N$, with generators $\ell_x$, $\ell_y$, and $f$. The classes of interferometric operators likewise form the group $\mathbb{Z}_N\times \mathbb{Z}_N\times \mathbb{Z}_N$, with generators $X$, $Y$, and $F$, where $X$ and $Y$ are cylindrical membrane operators along the $x$ and $y$-axes, and $F$ is a rigid wireframe operator. The precise form of the interferometric operators can be computed as a composition of Hamiltonian terms within a region encompassing $\mathcal{R}$. They exhibit the non-trivial statistical phases $S_{\ell_x,Y}= S_{\ell_y,X}=S_{f,F}=\omega$.

\subsection{Checkerboard model}
\label{sec:checkerboard}

\begin{figure}[htbp]
\centering
\includegraphics[width=.4\textwidth]{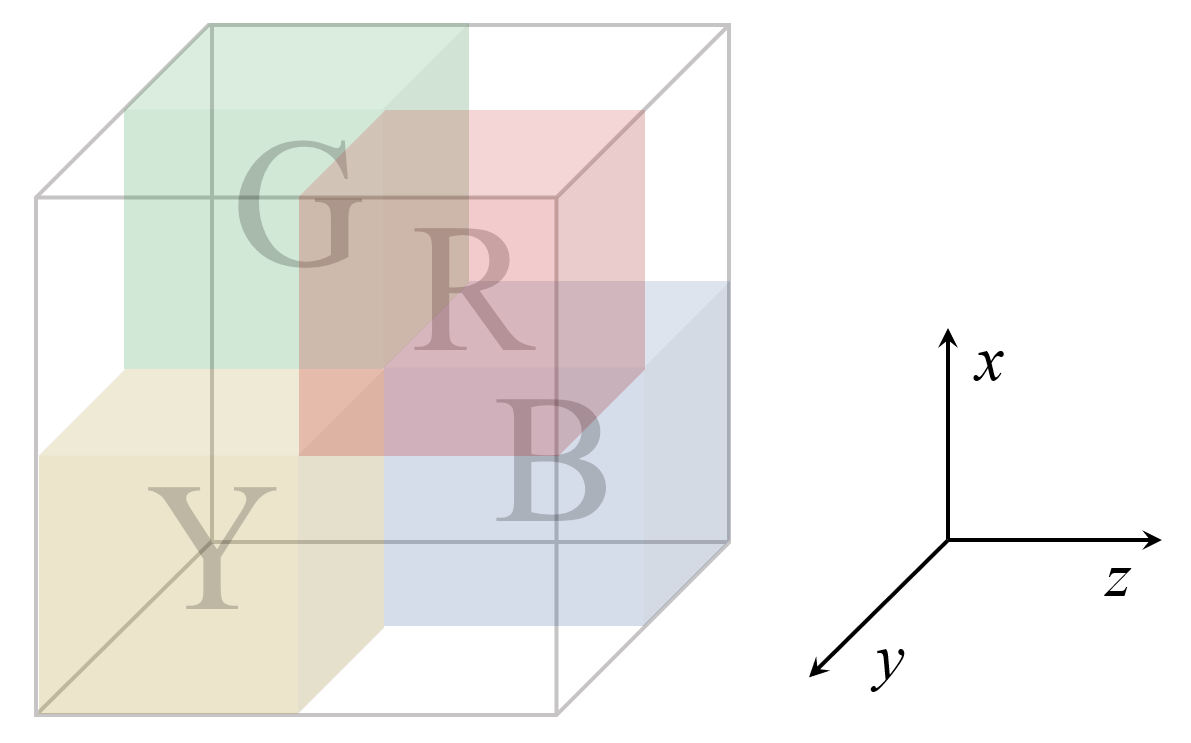}
\caption{The $A$ checkerboard sublattice is further subdivided into $R$, $G$, $B$, and $Y$ sublattices.}
\label{fig:RGBY}
\end{figure}

The checkerboard model, as introduced in Ref. \cite{Sagar16}, is a stabilizer code model defined on a cubic lattice with one qubit degree of freedom per site. The elementary cubes of the lattice are bipartitioned into $A$-$B$ checkerboard sublattices, and the Hamiltonian is defined as follows:
\begin{equation}
    H=-\sum_{c\in A}X_c-\sum_{c\in A}Z_c
\end{equation}
where $c\in A$ denotes the set of all cubes in sublattice $A$. The stabilizer generator $X_c$ ($Z_c$) is a product of Pauli $X$ ($Z$) operators over the vertices of cube $c$.

\begin{figure}[htbp]
\centering
\includegraphics[width=.9\textwidth]{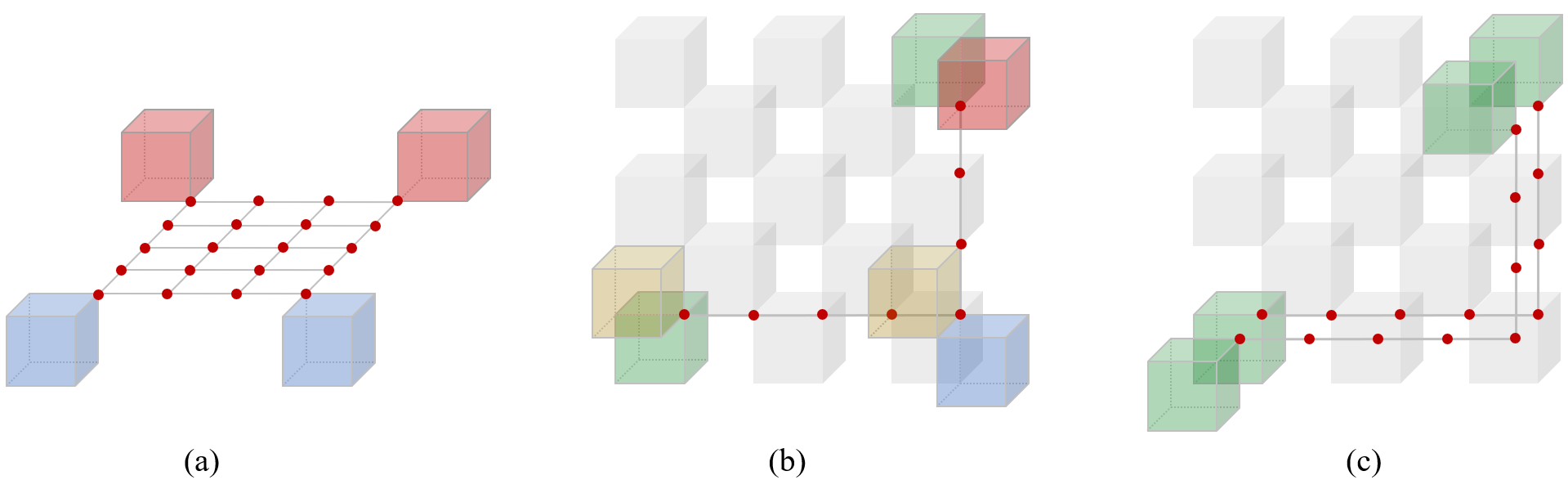}
\caption{Examples of (a) fracton excitations at the corners of membrane operators, (b) lineon excitations at the endpoints and corners of rigid string operators, and (c) planon excitations at the ends of flexible string operators in the checkerboard model. In all cases, the operators are products of Pauli $X$ or $Z$ over the red qubits.}
\label{fig:CBexcitations}
\end{figure}

To analyze the structure of fractional excitations in the model, it is convenient to regard a $2\times2\times2$ box as the elementary unit cell of the system, and to further subdivide the $A$ sublattice into $R$, $G$, $B$, and $Y$ sublattices, as pictured in \figref{fig:RGBY}. The model exhibits an `electric-magnetic' duality realized by Hadamard rotation which greatly simplifies the analysis. Let us first focus on the elementary electric excitations, which correspond to violations of individual $Z_c$ cube operators. They are immobile fractons that can only be created at the corners of membrane operators. Pairs of neighboring (i.e. sharing an edge) fracton excitations in differing sublattices (e.g. $R$ and $G$) are free to move along a line, and are thus lineons, whereas pairs of neighboring fractons in the same sublattice are planons with mobility in a 2D plane (see \figref{fig:CBexcitations}). Consequently, all the electric fractons in a single sublattice belong to the same QSS. However, fractons residing in different sublattices correspond to distinct quotient sectors, which are given labels $f^Z_R$, $f^Z_G$, $f^Z_B$, and $f^Z_Y$. Finally, because a composite of four adjacent electric fractons, one in each of the $R$, $G$, $B$, and $Y$ sublattices, is a local excitation (created by the action of a Pauli $X$ operator on a single qubit), each of these sectors is the result of fusion of the other three. In other words,
\begin{equation}
    f^Z_R\times f^Z_G\times f^Z_B\times f^Z_Y=1.
\end{equation}
Therefore, the electric excitations comprise 3 independent QSS. Likewise, there are 3 independent quotient sectors corresponding to magnetic quasiparticles, for a total of $2^6=64$ quotient sectors.

\begin{figure}[htbp]
\centering
\hspace{.03\textwidth}
$F^{X/Z}$ \hspace{.15\textwidth}
$M^{X/Z}_{BY}$ \hspace{.14\textwidth}
$M^{X/Z}_{GY}$ \hspace{.14\textwidth}
$M^{X/Z}_{GB}$ \hspace{.07\textwidth}
\includegraphics[width=0.9\textwidth]{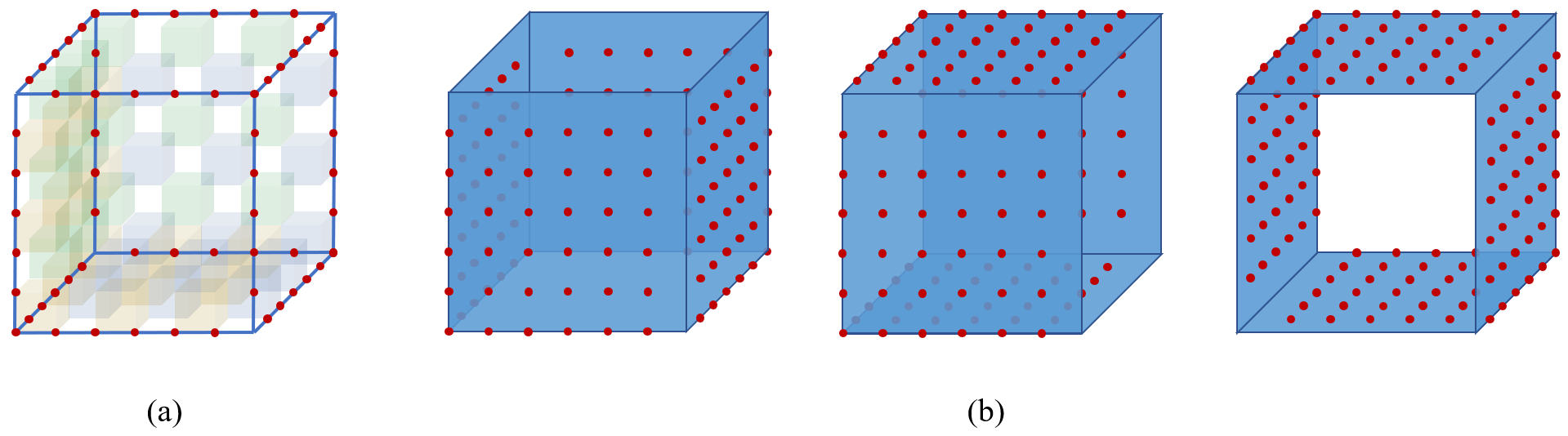}
\caption{Examples of (a) a wireframe operator and (b) membrane operators in the checkerboard model. The operators are tensor products of Pauli $X$ or $Z$ over the red qubits. Shaded cubes belong to the $A$ sublattice.}
\label{fig:CBoperators}
\end{figure}

Due to the self-duality, the interferometric operators of the checkerboard model may also be split according to whether they detect electric or magnetic excitations. Like the X-cube model, the checkerboard model has wireframe operators which correspond to processes in which lineons travel along the edges of the wireframe and fuse into the vacuum at the corners, as well as cylindrical membrane operators wrapping around one of three coordinate axes (for instance, as shown in \figref{fig:CBoperators}). The operators are tensor products of Pauli $X$ or $Z$ over the red qubits. The wireframe operators can be obtained as a product of all the $X_c$ or $Z_c$ cube operators inside the wireframe and are labeled as $F^X$ or $F^Z$, respectively. The membrane operators can be obtained as a product of all the cube operators in every other layer inside the overall cube. Depending on the orientation of the membrane operators, we label them as $M^X_{BY}$, $M^X_{GY}$, and $M^X_{GB}$ or as $M^Z_{BY}$, $M^Z_{GY}$, and $M^Z_{GB}$. The superscript denotes whether it is a tensor product of Pauli $X$ or $Z$, and the subscript specifies which layers of cubes; for instance, $M^Z_{BY}$ is a product of $Z_c$ over all $B$ and $Y$ cubes.

The structure of fractional excitations of the checkerboard model is \textit{identical} to that of two copies of the X-cube model. In other words, there is a mapping between quotient superselection sectors and interferometric operators of the two models which preserves the fusion rules and quasiparticle statistics, suggesting that the two models represent the same foliated fracton phase. In a separate work, we show that these models are in fact equivalent up to a generalized local unitary transformation\cite{Checkerboard}. The correspondence between non-trivial QSS and interferometric operators (IO) of the checkerboard model and two copies of the X-cube model is as follows:

\begin{center}
    \begin{tabular}{ |c|c||c|c| } 
     \hline
     Checkerboard QSS & QSS of 2 X-cube & Checkerboard IO & IO of 2 X-cube\\ \hline\hline
     $f^Z_R$ & $f^2$ & $F^Z$ & $F^2$ \\ \hline
     $f^Z_G$ & $\ell^1_x \times f^2$ & $M^Z_{BY}$ & $X^1$ \\ \hline
     $f^Z_Y$ & $\ell^1_y \times f^2$ & $M^Z_{GB}$ & $Y^1$ \\ \hline
     $f^Z_B$ & $\ell^1_z \times f^2$ & $M^Z_{GY}$ & $Z^1$ \\ \hline
     $f^Z_R\times f^Z_G$ & $\ell^1_x$ & $F^Z M^Z_{BY}$ & $F^2X^1$\\ \hline
     $f^Z_R\times f^Z_Y$ & $\ell^1_y$ & $F^Z M^Z_{GB}$ & $F^2Y^1$\\ \hline
     $f^Z_R\times f^Z_B$ & $\ell^1_z$ & $F^Z M^Z_{GY}$ & $F^2Z^1$\\ \hline\hline
     $f^X_R$ & $f^1$ & $F^X$ & $F^1$ \\ \hline
     $f^X_G$ & $\ell^2_x \times f^1$ & $M^X_{BY}$ & $X^2$ \\ \hline
     $f^X_Y$ & $\ell^2_y \times f^1$ & $M^X_{GB}$ & $Y^2$ \\ \hline
     $f^X_B$ & $\ell^2_z \times f^1$ & $M^X_{GY}$ & $Z^2$ \\ \hline
     $f^X_R\times f^X_G$ & $\ell^2_x$ & $F^X M^X_{BY}$ & $F^1X^2$ \\ \hline
     $f^X_R\times f^X_Y$ & $\ell^2_y$ & $F^X M^X_{GB}$ & $F^2Y^2$ \\ \hline
     $f^X_R\times f^X_B$ & $\ell^2_z$ & $F^X M^X_{GY}$ & $F^2Z^2$ \\ \hline
     \hline
    \end{tabular}
\end{center}
The superscripts in the X-cube columns indicate whether the sector or operator corresponds to the first or second X-cube copy. Note that there is an ambiguity in the correspondence due to the four-fold permutation symmetry of the $R$, $G$, $B$, and $Y$ sublattices; for example, we could have chosen $f^2$ ($f^1$) to correspond to $f^Z_G$ ($f^X_G$) instead of $f^Z_R$ ($f^X_R$), in which case $R$ and $G$ would be swapped in the above table.


\subsection{An anisotropic model with lineons and planons}
\label{sec:anisotropic}

\begin{figure}[htbp]
\centering
\includegraphics[width=.55\textwidth]{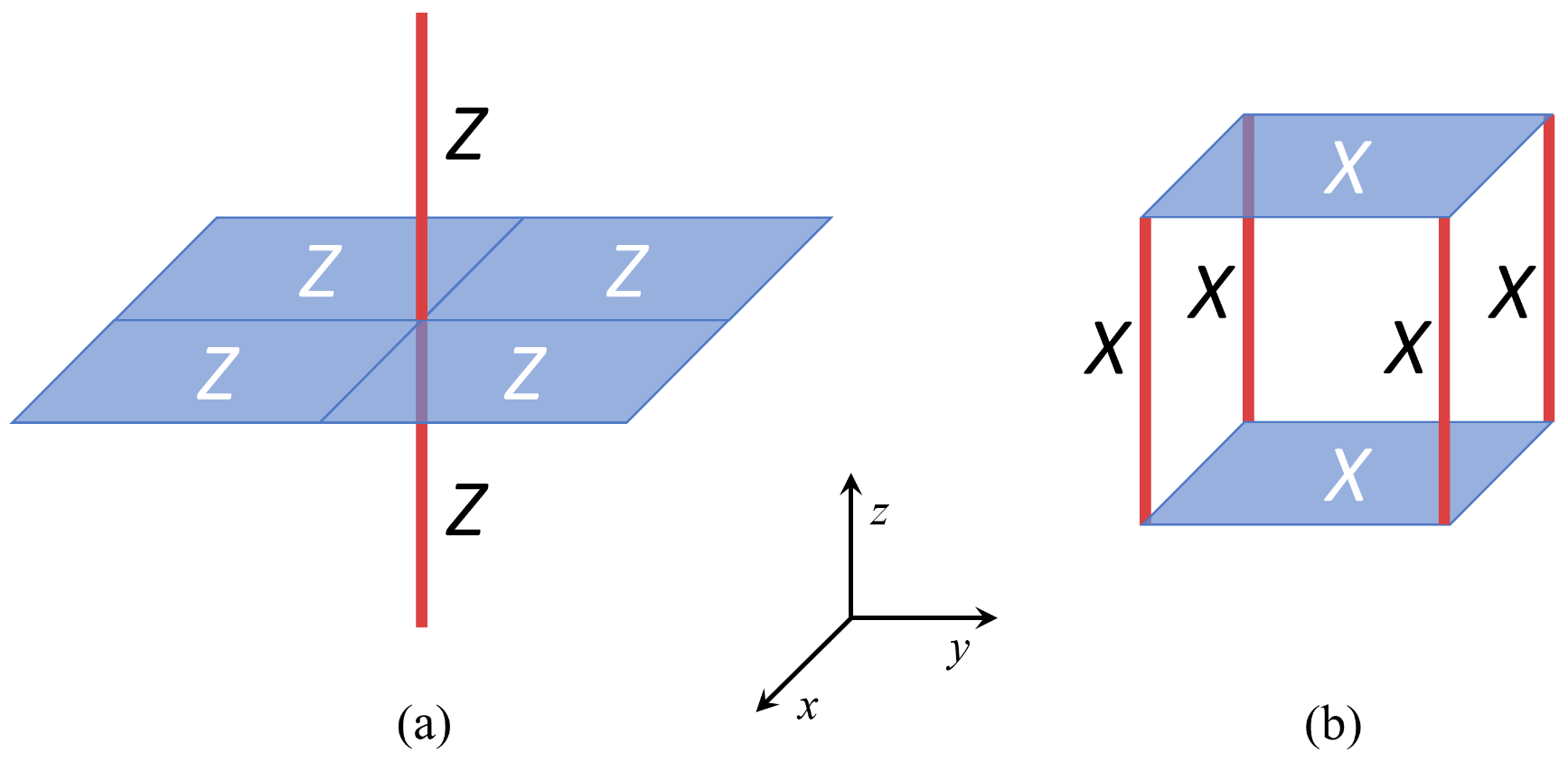}
\caption{The Hamiltonian terms of the anisotropic model. Qubits lie on the red edges and blue plaquettes.}
\label{fig:anisoH}
\end{figure}

In this section we discuss a novel stabilizer code Hamiltonian. In fact, it arises as a particular example of the polynomial formalism for translation-invariant stabilizer codes developed by Yoshida in Ref. \cite{YoshidaFractal}. The model is defined on a cubic lattice, with one qubit attached to each $z$-oriented link and one qubit attached to each $xy$ plaquette. The Hamiltonian takes the simple form
\begin{equation}
    H_\mathrm{aniso}=-\sum_v A_v -\sum_c B_c \label{eq:aniso}
\end{equation}
where $v$ runs over all vertices of the lattice, and $c$ runs over all elementary cubes. Here $A_v$ is defined as the product of Pauli $Z$ operators over the 4 plaquettes and 2 links adjacent to $v$, whereas $B_c$ is a product of Pauli $X$ operators over the 2 plaquettes and 4 links surrounding $c$ (as shown in \figref{fig:anisoH}).
The model exhibits a self-duality realized by duality of the underlying lattice composed with Hadamard rotation. It represents a foliated fracton phase with 2 foliations composed of toric code layers along the $xz$ and $yz$ planes. A fixed-point RG transformation for the model is discussed in \appref{app:anisotropicRG}. The model also admits a simple field theory description which is derived in \appref{app:anisotropic QFT}.

\begin{figure}[htbp]
\centering
\includegraphics[width=.6\textwidth]{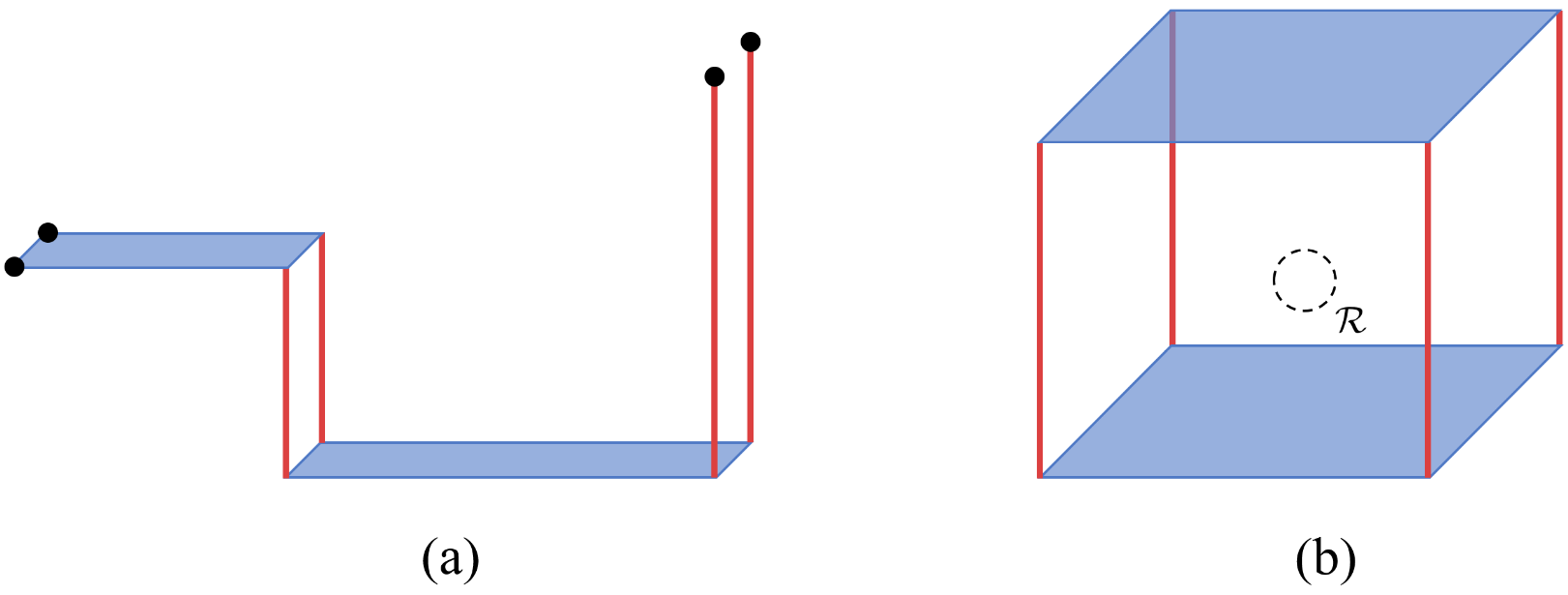}
\caption{(a) An open ribbon operator given by the tensor product of Pauli $X$ operators over the pictured qubits. Individual black dots represent lineons; the dipolar composites are planons. (b) An interferometric operator for the anisotropic model belonging to the $M$ class, given by the product of $X$ operators over the pictured qubits. The microscopic region $R$ lies at the center of the prism.}
\label{fig:anisofig2}
\end{figure}

There are two varieties of fractional excitations in this model: `electric' lineons, and `magnetic' lineons. The electric quasiparticles are created at the corners of membrane operators, which are tensor products of Pauli $X$ operators over the plaquettes in a region of a single $xy$ plane, and also at the ends of rigid string operators, which are tensor products of $X$ operators over the edges along a line segment oriented in the $z$ direction. These particles are individually only free to move in the $z$ direction, and are hence lineons. However, pairs of adjacent lineons are free to move in a 2D plane via the action of \textit{ribbon} operators, and are thus fractional planon excitations in their own right. Therefore, all electric lineon excitations belong to the same quotient superselection sector, which we label as $e$. An example of such a ribbon operator is depicted in \sfigref{fig:anisofig2}{a}.

Analogously, the magnetic excitations are created at the corners of membrane operators and the ends of string operators which are defined on the \textit{dual} lattice and are tensor products of Pauli $Z$ operators. These quasiparticles are likewise $z$ direction lineons, and pair to form dipolar planons. Thus the magnetic lineons represent a second non-trivial quotient superselection sector, labelled $m$. Finally, the composite of an electric and a magnetic lineon is a `dyonic' lineon which represents a non-trivial quotient sector labelled by $\epsilon$. The quotient sectors obey the simple fusion rules $e\times e=m\times m=1$, and $e\times m=\epsilon$.

The interferometric operators of this model correspond to compositions of Hamiltonian terms within some macroscopic region. Products of the cube terms, denoted as the class $M$ since such operators correspond to tunneling processes of magnetic lineons, detect the parity of electric lineons, whereas products of the vertex terms, denoted $E$, detect the parity of magnetic lineons. An example of an interferometric operator belonging to the $M$ class is shown in \sfigref{fig:anisofig2}{b}. Composite operators belonging to the class $\Sigma$ detect both types of lineons. The $S$ matrix, with respect to bases $\{1,e,m,\epsilon\}$ and $\{1,E,M,\Sigma\}$, is as follows:
\begin{equation}
    S=
    \begin{pmatrix}
        1 & 1  &   1 & 1  \\
        1 & 1  &  -1 & -1 \\
        1 & -1 & 1   & -1 \\
        1 & -1 & -1  & 1  \\
    \end{pmatrix}.
\end{equation}

\section{Mapping the semionic X-cube model to the X-cube model}
\label{sec:mapping}

As discussed in section \ref{sec:SXc}, the semionic X-cube model has the same quotient superselection sectors and interferometric statistics as the X-cube model, indicating that they may belong to the same foliated fracton phase. In this section, we show that this is indeed the case by presenting an explicit mapping between the two. Note that, as discussed in Ref.~\cite{MaLayers}, the two models appear to be very different because in the X-cube model string operators of lineons always commute with each other, while in the semionic X-cube model string operators of lineons may anti-commute with each other (if they lie in orthogonal directions and intersect one another). However, as we see below, this difference is merely superficial and can be removed by considering the general equivalence relation used to define foliated fracton phases. In fact, to map between the two models, we must first add stacks of 2D double semion layers in the $xy$, $yz$, and $zx$ planes to both models before applying local unitary transformations. In the presence of such layers, the two models become equivalent. One way to see this equivalence is to realize that with these layers, we can bind the 2D semions from the layers to the lineons in the model, hence changing the string operators of the lineons from commuting to anti-commuting or vice versa. Therefore, in the presence of the double semion layers, the two models are no longer distinct.

\begin{figure}[htbp]
\centering
\includegraphics[width=0.6\textwidth]{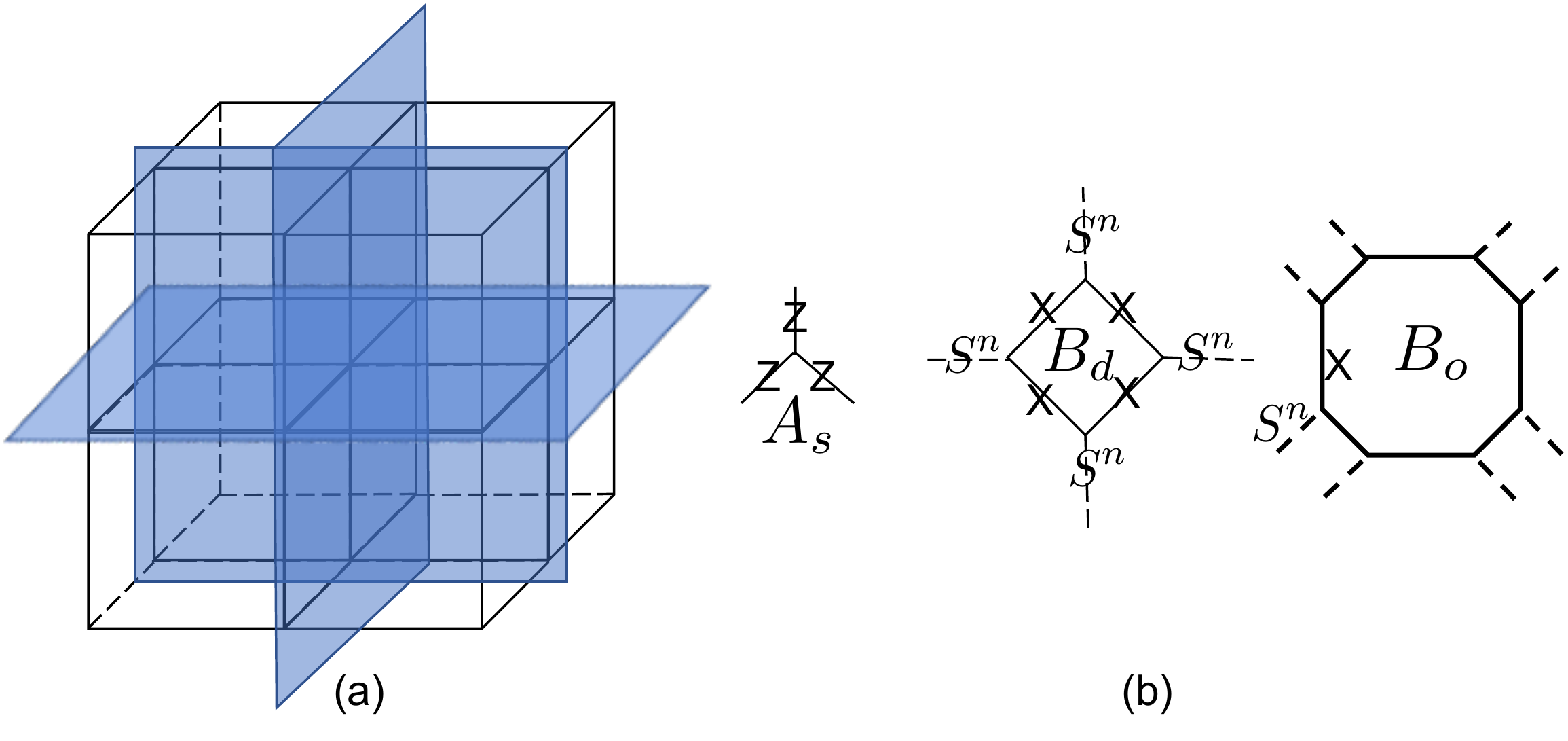}
\caption{(a) Inserting stacks of double semion layers (blue planes) along the $xy$, $yz$ and $zx$ planes into the X-cube and semionic X-cube models. The layers overlap with the $xy$, $yz$, and $zx$ planes of the decorated cubic lattice. For clarity, the decoration (which is shown in \figref{fig:triv_stack}) in the cubic lattice is not shown and only one layer is shown for each stack.
(b) The vertex and plaquette Hamiltonian terms of the 2D toric code model ($n=0$) and 2D double semion model ($n=1$) on the decorated square lattice. In the $B_o$ term, there is one $X$ operator on each solid edge and one $S^n$ operator on each dashed edge. For clarity, only two of these operators in the right-most figure are shown.}
\label{fig:mapping}
\end{figure}

The mapping goes as follows. We add to the decorated cubic lattice three stacks of double semion layers in the $xy$, $yz$, and $zx$ planes, as shown in \sfigref{fig:mapping}{a}. The double semion models are defined on decorated square lattices as shown in \sfigref{fig:triv_stack}{a}. With this addition, the two models take the form
\begin{equation}
H_n' = -\sum_v A_v -\sum_p B^{(n)}_p - \sum_c C^{(n)}_c - \sum_L \left(\sum_{s\in L} A_s + \sum_{d \in L} B^{(1)}_d + \sum_{o\in L} B^{(1)}_o\right),
\end{equation}
where $n=0$ for the X-cube model and $n=1$ for the semionic X-cube model. The $A_v$, $B^{(n)}_p$, $C^{(n)}_c$ terms are given in \sfigref{fig:triv_stack}{c-e}. The $A_s$ (vertex), $B^{(1)}_d$ (diamond plaquette), $B^{(1)}_o$ (octagon plaquette) terms belong to each double semion layer labeled by $L$ and take the form as shown in \sfigref{fig:mapping}{b}. The difference between the two models lies in the $B_p$, $C_c$ terms while all other terms are the same. Each $B_p$ term overlaps with one $B^{1}_d$ term in the double semion layers while each side surface of the $C_c$ term overlaps with one $B^{(1)}_o$ term in these layers. Therefore, to map between the two models, it suffices to show that the combination of the $B_p$, $B^{(1)}_d$ terms and the combination of the $C_c$, $B^{(1)}_o$ terms can be mapped from one model to the other without affecting the other terms.

To establish this mapping, first we consider a 2D problem of mapping from one 2D toric code model plus one 2D double semion model to two copies of the 2D double semion model. The Hamiltonian of the first system is given by
\begin{equation}
H_a = \sum_{s_1\in L_1} A_{s_1} + \sum_{d_1 \in L_1} B^{(0)}_{d_1} + \sum_{o_1 \in L_1} B^{(0)}_{o_1} + \sum_{s_2 \in L_2} A_{s_2} + \sum_{d_2 \in L_2} B^{(1)}_{d_2} + \sum_{o_2 \in L_2} B^{(1)}_{o_2}. \label{eq:Ha}
\end{equation}
Here $L_1$ and $L_2$ are two separate layers. The Hamiltonian for the second system is given by
\begin{equation}
H_b = \sum_{s_1\in L_1} A_{s_1} + \sum_{d_1 \in L_1} B^{(1)}_{d_1} + \sum_{o_1 \in L_1} B^{(1)}_{o_1} + \sum_{s_2 \in L_2} A_{s_2} + \sum_{d_2 \in L_2} B^{(1)}_{d_2} + \sum_{o_2 \in L_2} B^{(1)}_{o_2}. \label{eq:Hb}
\end{equation}
It is possible to map between these two models with local unitary transformations because they have the same topological order. This can be seen by observing that both models represent $\mathbb{Z}_2\times \mathbb{Z}_2$ gauge theories containing two independent gauge charges $c_1$, $c_2$ and two independent gauge fluxes $f_1$, $f_2$. The statistics of the two models are similar in the following ways: 
\begin{align}
t_{c_1}&=1,&
t_{c_2}&=1,& s_{c_1f_1}&=-1,& s_{c_2,f_2}&=-1, \nonumber\\ s_{c_1,c_2}&=1,& s_{f_1,f_2}&=1,& s_{c_1,f_2}&=1,& s_{c_2,f_1}&=1
\end{align}
where $t$ denotes topological spin, and $s$ denotes the braiding statistics. The two models are different in the topological spin for the two fluxes. In model $a$:
\begin{align}
t_{f_1} &= 1,& t_{f_2}&=i
\end{align}
In model $b$:
\begin{align}
t_{f_1} &= i, &t_{f_2}&=i
\end{align}
But this difference is only superficial because we can reorganize the quasiparticles of model $a$ so that they have the same statistics as model $b$. In particular, if we redefine the quasiparticles in model $a$ as
\begin{align}
c_1'&=c_1,& c_2'&=c_1c_2,& f_1'&=c_1f_1c_2f_2,& f_2'&=f_2,
\end{align}
then they have the same statistics as model $b$. Therefore, there exists a local unitary transformation mapping the ground state of model $a$ to the ground state of model $b$. At the same time, it maps $c_1$ in model $a$ to $c_1$ in model $b$, $c_1c_2$ to $c_2$, $c_1f_1c_2f_2$ to $f_1$, $f_2$ to $f_2$. Correspondingly, it maps the Hamiltonian terms, which are also loop operators of the quasiparticles, as follows
\begin{align}
A_{s_1} &\to A_{s_1},& A_{s_1}A_{s_2} &\to A_{s_2},& B^{(0)}_{d_1}B^{(1)}_{d_2} &\to B^{(1)}_{d_1}, \nonumber\\ B^{(1)}_{d_2} &\to B^{(1)}_{d_2},& B^{(0)}_{o_1}B^{(1)}_{o_2} &\to B^{(1)}_{o_1},& B^{(1)}_{o_2} &\to B^{(1)}_{o_2}
\label{map2D}
\end{align}

More explicitly, the local unitary transformation involves a controlled-$X$ operator from every qubit in $L_1$ to its counterpart qubit in $L_2$, followed by a unitary on the six qubits around each pair of corresponding vertices in the two layers. The unitary is diagonal in the computational basis of the six qubits $U=\sum_{a,b,c} \alpha(a,b,c)|a,b,c\rangle\langle a,b,c|$. Here $a,b,c=0,1,2,3$ label the $(0,0),(0,1),(1,0),(1,1)$ state of each pair of corresponding qubits in the two layers. $\alpha(a,b,c)$ is given as follows:
\begin{gather}
\begin{gathered}
\begin{aligned}
\alpha(0,0,0)&=1,& \alpha(1,3,2)&=-1,
\end{aligned} \\
\alpha(1,1,0)=\alpha(2,2,0)=\alpha(3,3,0)=\alpha(1,2,3)=i. \end{gathered}
\end{gather}
$\alpha$ is invariant under cyclic permutations of $a,b,c$. All other terms of $\alpha$ are $1$.

The equivalence between $H_a$ and $H_b$ [Eqns.~\eqref{eq:Ha} and \eqref{eq:Hb}] can also be understood in the $K$-matrix formalism. These models have a Chern-Simons description \cite{WenChernSimons} given by the following Lagrangian and respective $K$-matrices, where $a^I_\mu$ is a compact gauge field:
\begin{align}
  \mathcal{L} &= \frac{1}{4\pi} K_{IJ} \epsilon^{\mu\nu\rho} a^I_\mu \partial_\nu a^J_\rho \\[.2cm]
  K_a &= \begin{pmatrix}0&2&0&0\\2&0&0&0\\0&0&2&0\\0&0&0&-2\end{pmatrix} \\[.2cm]
  K_b &= \begin{pmatrix}2&0&0&0\\0&-2&0&0\\0&0&2&0\\0&0&0&-2\end{pmatrix}.
\end{align}
However, the $K$-matrix for a Chern-Simons Lagrangian is not unique.
A change of variables can be performed by transforming the gauge fields according to $a^I_\mu \to a'^{I'}_\mu = W^{I'}_I a^{I}_\mu$ where $W \in \text{GL}(n,\mathbb{Z})$ is an $n \times n$ matrix (with $n=4$ for $K_a$ and $K_b$) with integer coefficients and determinant equal to 1.
This transformation can then be absorbed into the $K$-matrix via $K_{IJ} \to K'_{I'J'} = W^I_{I'} K_{IJ} W^J_{J'} = (W^T K W)_{I'J'}$.
The following matrix $W$ can then be used to relate the above two $K$-matrices:
\begin{equation}
  W = \begin{pmatrix}0&-1&+1&0\\+1&+1&0&0\\-1&-1&+1&0\\0&0&0&+1\end{pmatrix}. 
\end{equation}
In particular, $K_b = W^T K_a W$. This shows that $H_a$ and $H_b$ represent the same phase.

Using such a local unitary transformation it is possible to map between the X-cube model and the semionic X-cube model when both are augmented with double semion layers. This is because we can apply the local unitary transformation between a double semion layer and the Hamiltonian terms in the overlapping layer of the X-cube model and map from
\begin{align}
A_v &\to A_v, &A_vA_s &\to A_s,& B_p^{(0)}B_d^{(1)} &\to B_p^{(1)}, \nonumber\\
B_d^{(1)} &\to B_d^{(1)},& C_c^{(0)}\left(B_o^{(1)}\right)^{\otimes 6} &\to C_c^{(1)}, &B_o^{(1)} &\to B_o^{(1)}.
\end{align}
Of course, the X-cube and semionic X-cube models are different from simple decoupled stacks of toric code and double semion models, so there are some subtleties involved in applying the mapping in Eq.~\ref{map2D}. In particular, the X-cube and semionic X-cube models are `coupled' toric codes and double semions such that loop configurations on the side surfaces of the same cube should exist at the same time. While performing the mapping, care must be taken that this constraint is not violated. Indeed this is the case because in mapping between $H_a$ and $H_b$, $f_1$ maps to $f_1f_2$ and $f_1f_2$ to $f_1$, therefore the loop configuration in the first layer is always preserved. Applying the same mapping to the X-cube or semionic X-cube models together with double semion stacks, the loop configurations in these models are also always preserved.
 
In this way, it is possible to map the ground state of the X-cube model to the ground state of the semionic X-cube model after inserting three stacks of double semion layers in the $xy$, $yz$, $zx$ directions respectively. Ref.~\cite{YouSSPT} discussed an ungauged version of the semionic X-cube model as a symmetry protected topological (SPT) phase with subsystem symmetry. Using a similar transformation, one can show that the ungauged model is equivalent (with the addition of 2D $\mathbb{Z}_2$ SPT layers) to a `weak' subsystem SPT model which is a stack of 2D SPTs. 

\section{Loop excitations}

\label{sec:loop}

In three dimensions, gapped topological phases harbor fractional loop-like excitations in addition to point-like particles. Moreover, these loop-shaped excitations may exhibit non-trivial braiding statistics with point particles, as well as from three-loop braiding processes in which two loops are wound around one another while simultaneously linked to a third loop.\cite{Wang2014a,Jiang2014,Wang2015}
In conventional 3D topological phases (discrete gauge theories), the set of fractional particle and loop excitations, the braiding statistics between particles and loops, along with the three-loop braiding statistics, fully characterize the topological order (see, for example, Refs. \cite{WangLevin15,Cheng2018}). In this section, we demonstrate that the framework developed in the prior sections can be extended to accommodate the universal data pertaining to loop excitations of conventional topological orders. This is of note because conventional 3D topological orders are themselves a subset of the foliated fracton orders (with trivial foliation structure). Since the notions of ordinary superselection sector and quotient superselection sector coincide for these phases, there is no need to distinguish between them here.

The conventional notion of \textit{superselection sector} does not capture loop-like excitations of gapped phases, because loop excitations contained in a ball-shaped region $\mathcal{R}$ can be shrunk to a point and annihilated via the action of a local operator with support in $\mathcal{R}$. However, it is possible to incorporate a description of these excitations by modifying the topology of the region. Instead of a ball-shaped region $\mathcal{R}$, consider a region $\mathcal{S}$ with the topology of a solid torus. We assume that the diameter of $\mathcal{S}$ (but not necessarily the thickness) is large compared to the correlation length of the gapped medium. The superselection sectors defined with reference to such a region $\mathcal{S}$ include the original sectors corresponding to fractional point particles, as well as new sectors which correspond to fractional loop excitations. For example, for the 3+1D $\mathbb{Z}_2$ gauge theory, there are four such superselection sectors: the vacuum, an electric point charge $e$, a magnetic flux loop $m$, and a dyonic loop $\epsilon$, which is a composite of $e$ and $m$ excitations and carries both charge and flux. The interferometric operators for these sectors correspond to processes in which a charge is wound around a flux loop, or in which a flux loop is nucleated from the vacuum, stretched and pulled around a charge, and annihilated into the vacuum on the other side. It is also possible to capture the notion of three-loop braiding by extending the notion of superselection sector to regions with the topology of Hopf-linked solid tori. We will not elaborate further here.

\section{Discussion}

\label{sec:discussion}

In this paper, we have proposed a way to characterize fractional excitations in fracton models that reflects the universal properties of the underlying foliated fracton order. A foliated fracton phase is defined to be the equivalence class of 3D gapped fracton models up to the addition of 2D gapped topological layers and adiabatic deformation. Correspondingly, we propose to characterize fractional excitations in fracton models by modding out the contributions from the 2D layers. We define a \textit{quotient superselection sector} (QSS), coarsening the notion of superselection sectors, as an equivalence class of point excitations that can be related to one another by adding or removing both local excitations and 2D quasiparticles. Moreover, we define their `statistics' in an interferometric way that is indifferent to statistics arising from the exchange or braiding of 2D quasi-particles in the system. Using this approach, we can characterize the universal features of fractional excitations in a foliated fracton phase using a finite data set and compare this structure between models. The examples we studied fall into three classes, as summarized in the following table. (From our preliminary studies, the Chamon model \cite{ChamonModel,ChamonModel2} belongs to the class of X-cube model with 4 foliations. Details about this model will be presented in future work.)

\begin{center}
 \begin{tabular}{| c | c | c | c |} 
 \hline
 \multirow{2}{*}{Class} & Independent & Independent & \multirow{2}{*}{Models} \\
 & Fracton QSS & Lineon QSS & \\
 \hline\hline
 \multirow{3}{*}{\begin{tabular}{c}X-cube \\ 3-foliation\end{tabular}} & \multirow{3}{*}{1} & \multirow{3}{*}{2} & X-cube, semionic X-cube, \\ 
 & & & $\mathbb{Z}_N$ X-cube, \\
 & & & Checkerboard (2 copies) \\
 \hline
 \multirow{3}{*}{\begin{tabular}{c}X-cube \\ 4-foliation\end{tabular}}
   & \multirow{3}{*}{1} & \multirow{3}{*}{3} & Kagome X-cube \\
 & & & Hyperkagome X-cube \\
 & & & Chamon model \\
 \hline
 Anisotropic & \multirow{2}{*}{0} & \multirow{2}{*}{2} & \multirow{2}{*}{Anisotropic model} \\
 2-foliation & & & \\
 \hline
\end{tabular}
\label{table}
\end{center}

Within each class, the quasiparticle statistics given by interferometric detection also take the same form. Of course, this is not meant to be a complete list. It will be interesting to study the fractional excitations in the Majorana checkerboard model \cite{VijayFracton}, the non-abelian fracton models \cite{VijayNonabelian}, the twisted fracton models \cite{SongTwistedFracton}, the cage-net models \cite{CageNet}, and so forth. Compared to the systematic characterization of 2D fractional excitations in terms of unitary modular tensor categories, our understanding of fractional excitations in 3D fracton models is very limited. To achieve a more complete understanding, we must collect more data and determine what types of quotient superselection sectors can exist and what kinds of quasiparticle statistics are possible.

\section*{Acknowledgement}

We are grateful for inspiring discussions with David Aasen, Han Ma, and Michael Hermele. W.S. and X.C. are supported by the National Science Foundation under award number DMR-1654340 and the Institute for Quantum Information and Matter
at Caltech. X.C. is also supported by the Alfred P. Sloan research fellowship and the Walter Burke Institute for Theoretical Physics at Caltech.
K.S. is grateful for support from the NSERC of Canada, the Center for Quantum Materials at the University of Toronto, and the Walter Burke Institute for Theoretical Physics at Caltech.

\appendix

\section{RG transformation for the anisotropic model}
\label{app:anisotropicRG}

\begin{figure}
    \centering
    \begin{tikzpicture}
    
    \pgfmathsetmacro{\l}{1.3}
    
        \draw[arrows={->}] (-2.1*\l,-\l,0) -- ++(\l/2,0,0);
        \draw[arrows={->}] (-2.1*\l,-\l,0) -- ++(0,\l/2,0);
        \draw[arrows={->}] (-2.1*\l,-\l,0) -- ++(0,0,\l/1.5);
        
        \draw (-2.0*\l,-.6*\l,0) node[fill=none] {\scriptsize $z$};
        \draw (-1.7*\l,-1.15*\l,0) node[fill=none] {\scriptsize $y$};
        \draw (-2.2*\l,-1.3*\l,0) node[fill=none] {\scriptsize $x$};
    
        \draw[line width=.6]
            (-\l,0,0) --++ (0,0,-\l)
            (0,-\l,0) --++ (0,0,-\l)
            (0,0,0) --++ (0,0,-\l)
            (0,0,0) -- ++(-\l,0,0) -- ++(0,-\l,0) -- ++(\l,0,0) -- cycle
            (0,-\l,-\l) -- ++(0,\l,0) -- ++(-\l,0,0)
        ;
        \draw[dotted,line width=.6]
            (-\l,-\l,0) --++ (0,0,-\l)
            (-\l,-\l,-\l) --++ (0,\l,0)
            (-\l,-\l,-\l) --++ (\l,0,0)
        ;
        \draw[line width=.6]
            (\l,0,0) --++ (0,0,-\l)
            (2*\l,-\l,0) --++ (0,0,-\l)
            (2*\l,0,0) --++ (0,0,-\l)
            (2*\l,0,0) -- ++(-\l,0,0) -- ++(0,-\l,0) -- ++(\l,0,0) -- cycle
            (2*\l,-\l,-\l) -- ++(0,\l,0) -- ++(-\l,0,0)
        ;
        \draw[dotted,line width=.6]
            (\l,-\l,0) --++ (0,0,-\l)
            (\l,-\l,-\l) --++ (0,\l,0)
            (\l,-\l,-\l) --++ (\l,0,0)
        ;
        
        \pgfmathsetmacro{\d}{-.15*\l}
        
        \draw[line width=.6]
            (\d+5*\l,0,0) -- ++(-\l,0,0) -- ++(0,-\l,0) -- ++(\l,0,0) -- cycle
        ;
        \draw[line width=.6]
            (\d+6.8*\l,0,0) -- ++(-\l,0,0) -- ++(0,-\l,0) -- ++(\l,0,0) -- cycle
        ;
        
        \draw(-\l,-\l,0) node[fill=white] {\scriptsize $ZZ$};
        \draw(-\l,0,0) node[fill=white] {\scriptsize $ZI$};
        \draw(-\l,-\l,-\l) node[fill=white] {\scriptsize $IZ$};
        \draw(0,-\l,-\l) node[fill=white] {\scriptsize $IZ$};
        \draw(0,-\l,0) node[fill=white] {\scriptsize $IZ$};
        
        \draw(2*\l,0,-\l) node[fill=white] {\scriptsize $XX$};
        \draw(1*\l,0,-\l) node[fill=white] {\scriptsize $XI$};
        \draw(1*\l,0,0) node[fill=white] {\scriptsize $XI$};
        \draw(2*\l,0,0) node[fill=white] {\scriptsize $XI$};
        \draw(2*\l,-\l,-\l) node[fill=white] {\scriptsize $IX$};
        
        \draw(\d+4*\l,-\l,0) node[fill=white] {\scriptsize $ZZ$};
        \draw(\d+4*\l,0,0) node[fill=white] {\scriptsize $ZI$};
        \draw(\d+5*\l,-\l,0) node[fill=white] {\scriptsize $IZ$};
        
        \draw(\d+6.8*\l,0,0) node[fill=white] {\scriptsize $XX$};
        \draw(\d+5.8*\l,0,0) node[fill=white] {\scriptsize $XI$};
        \draw(\d+6.8*\l,-\l,0) node[fill=white] {\scriptsize $IX$};
        
        \pgfmathsetmacro{\f}{-2.9*\l}
        
        \draw(.5*\l,-1.5*\l,0) node[fill=white] {\scriptsize (a)};
        \draw(\d+5.4*\l,-1.5*\l,0) node[fill=white] {\scriptsize (b)};
        \draw(.5*\l,\f-.5*\l,0) node[fill=white] {\scriptsize (c)};
        
        \draw[line width=.6]
            (0,\f,-\l) -- ++(-\l,0,0) -- ++(0,0,\l) -- ++(\l,0,0) -- cycle
        ;
        \draw[line width=.6]
            (2*\l,\f,-\l) -- ++(-\l,0,0) -- ++(0,0,\l) -- ++(\l,0,0) -- cycle
        ;
        
        \draw(-\l,\f,-\l) node[fill=white] {\scriptsize $IZ$};
        \draw(-\l,\f,0) node[fill=white] {\scriptsize $ZZ$};
        \draw(0,\f,0) node[fill=white] {\scriptsize $IZ$};
        \draw(0,\f,-\l) node[fill=white] {\scriptsize $IZ$};
        
        \draw(\l,\f,-\l) node[fill=white] {\scriptsize $XI$};
        \draw(\l,\f,0) node[fill=white] {\scriptsize $XI$};
        \draw(2*\l,\f,0) node[fill=white] {\scriptsize $XI$};
        \draw(2*\l,\f,-\l) node[fill=white] {\scriptsize $XX$};

    \end{tikzpicture}
    \caption{Terms of (a) the anisotropic model Hamiltonian $H_\mathrm{aniso}$, (b) the 2D toric code Hamiltonian $H_\mathrm{TC}$, which acts on qubits in the $x=x_0$ layer, and (c) the Hamiltonian $H_0$, which acts on the $z=z_0$ layer.}
    \label{fig:anisoH2}
\end{figure}
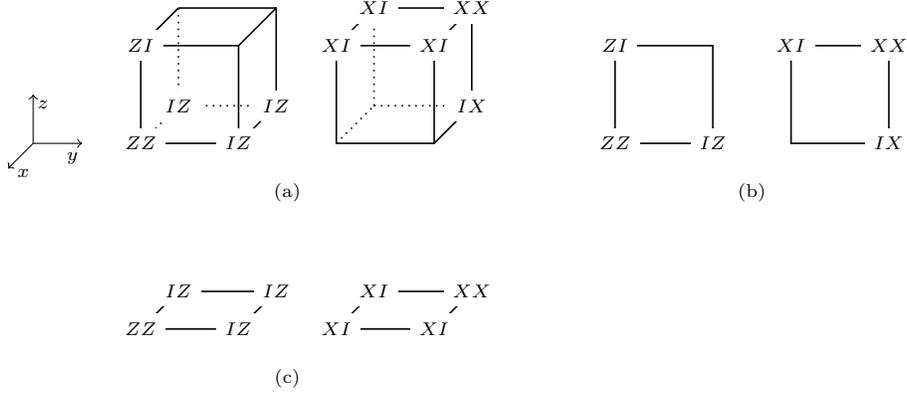

In this appendix, we discuss the renormalization group transformation for the anisotropic model introduced in \secref{sec:anisotropic}. The procedure utilizes 2D toric code resource states to grow the sytsem size in the $x$ or $y$ directions, and product state ancilla degrees of freedom to grow the system in the $z$ direction. Hence the model has foliated fracton order with 2 underlying foliations. To describe these transformations, it is convenient to re-arrange the qubits so that two qubits lie at each vertex of a cubic lattice. They may then be referred to by labels $(x,y,z,\alpha=1,2)$. In this geometry the Hamiltonian terms take the form pictured in \sfigref{fig:anisoH2}{a}.

To disentangle the layer $x=x_0$ from the rest of the system, we act with the local unitary operator
\begin{equation}
    S=\prod_{y,z} \mathrm{CNOT}_{(x_0-1,y,z,2),(x_0,y,z,2)}
    \mathrm{CNOT}_{(x_0,y,z,1),(x_0+1,y,z,1)},
\end{equation}
which satisfies $SH_\mathrm{aniso}S^\dagger\cong H'_\mathrm{aniso}+H_\mathrm{TC}$. Here, $H_\mathrm{aniso}$ is the original Hamiltonian for the anisotropic model, $H'_\mathrm{aniso}$ is the Hamiltonian for the model with the $x=x_0$ layer missing, and $H_\mathrm{TC}$ is the toric code Hamiltonian on the $x=x_0$ layer, whose stabilizer terms are depicted in \sfigref{fig:anisoH2}{b}. An analogous transformation can be used to disentangle 2D toric code layers along $xz$ planes. In order to grow the system size, this procedure is simply reversed: 2D toric code resource states are added to the 3D system then sewn into the bulk by the circuit $S$ (note that $S=S^{-1}$).

On the other hand, to disentangle the $z=z_0$ layer from the other system, we perform the operation
\begin{equation}
    S=\prod_{x,y} \mathrm{CNOT}_{(x_0-1,y,z,2),(x_0,y,z,2)}
    \mathrm{CNOT}_{(x_0,y,z,1),(x_0+1,y,z,1)},
\end{equation}
which acts as $SH_\mathrm{aniso}S^\dagger\cong H'_\mathrm{aniso}+H_0$. The decoupled Hamiltonian $H_0$, acting on the $z=z_0$ layer, is a sum of terms depicted in \sfigref{fig:anisoH2}{c}. This Hamiltonian has trivial topological order with a product state ground state. Therefore, the anisotropic model has an underlying foliation structure composed of 2 foliations of 2D topologically ordered gapped states parallel to the $xz$ and $yz$ planes.

\section{Field theory of the anisotropic model}
\label{app:anisotropic QFT}

In this appendix we derive a quantum field theory (QFT) for the anisotropic model introduced in \secref{sec:anisotropic}.
The QFT and its derivation are analogous to that of the X-cube model in \rcite{Slagle17QFT}.

\begin{figure}[htbp]
\centering
\includegraphics[width=.55\textwidth]{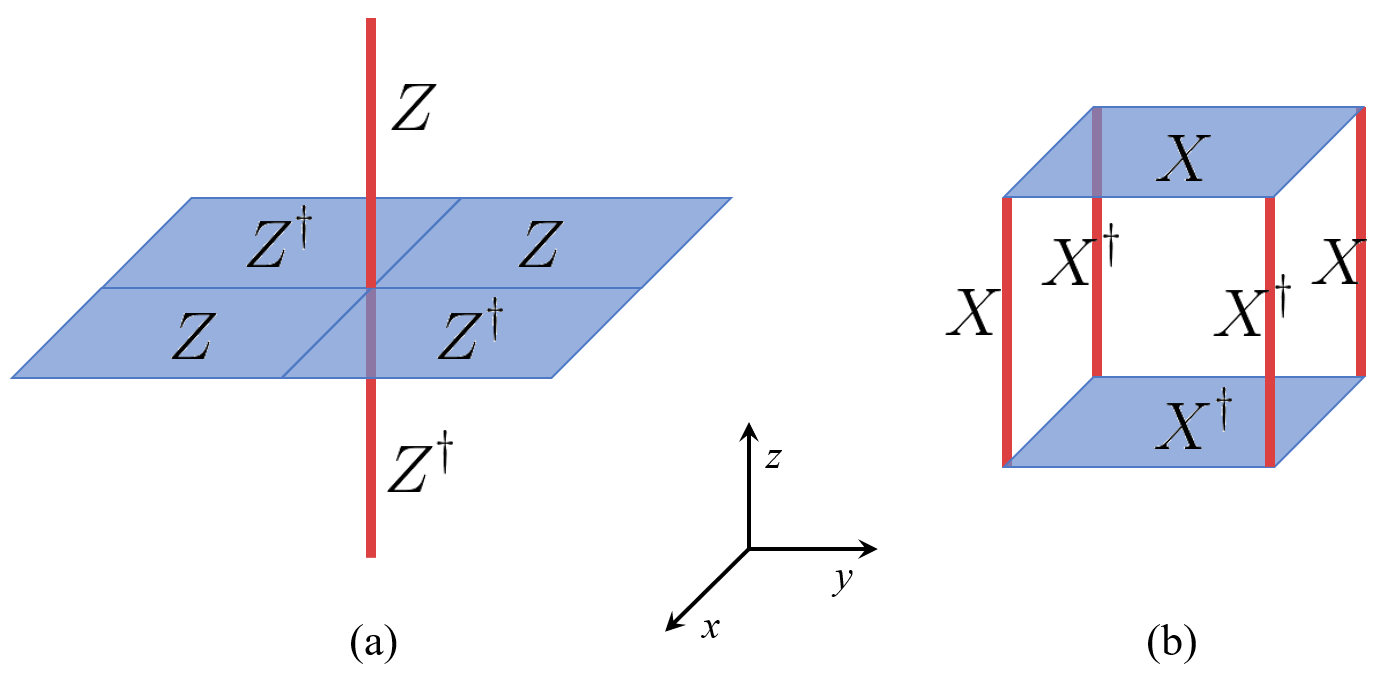}
\caption{
The Hamiltonian terms of the $\mathbb{Z}_N$ anisotropic model.
Rotor degrees of freedom lie on the red edges and blue plaquettes.}
\label{fig:ZNanisoH}
\end{figure}

We will consider the $\mathbb{Z}_N$ generalization of the anisotropic model.
The stabilizer terms in the Hamiltonian [\eqnref{eq:aniso}] are shown in \figref{fig:ZNanisoH}.
The $\mathbb{Z}_N$ rotor degrees of freedom ($X$ and $Z$) and defined the same way as in \secref{sec:ZNXc}.

In order to connect the lattice model to a field theory, we rewrite the lattice operators as exponents of fields ($A_b$ and $B_b$ with $b=1,2$):
\begin{align}
\begin{split}
 Z_{p(\mathbf{x})}(t) &\sim \exp\left( \ii \int_{x-a/2}^{x+a/2} \dd x' \int_{y-a/2}^{y+a/2} \dd y' A_1(t,x',y',z) \right) \\
 Z_{e(\mathbf{x})}(t) &\sim \exp\left( \ii \int_{z-a/2}^{z+a/2} \dd z' A_2(t,x,y,z') \right). \label{eq:operatorsAsFields}
\end{split}
\end{align}
The $X$ operators are related to the $B$ fields by replacing $Z\to X$ and $A\to B$ above.
$Z_{p(\mathbf{x})}(t)$ denotes a $Z(t)$ operator (in the Heisenberg representation) at the $xy$ plane plaquette $p(\mathbf{x})$, which is centered at $\mathbf{x}$,
  while $Z_{e(\mathbf{x})}(t)$ denotes a $Z(t)$ operator at the $z$ axis edge $e(\mathbf{x})$.
If $a$ is the cubic lattice spacing, then the $Z$ (or $X$) operators are related to the exponent of small integrals of the gauge field $A$ (or $B$) over plaquettes or lines of length $a$.
The $A$ and $B$ fields should not be confused with the stabilizer operators $A_v$ and $B_c$ appearing in the Hamiltonian [\eqnref{eq:aniso}].

The field theory is then derived by first rewriting the stabilizer terms in the Hamiltonian in terms of the field variables.
To do this, we express the stabilizers,
  \sfigref{fig:ZNanisoH}{a} and \hyperref[fig:ZNanisoH]{(b)},
  as exponents of current densities, $e^{\ii I^0}$ and $e^{\ii J^0}$, respectively.
The current densities are
\begin{align}
\begin{split}
I^0 &= \frac{N}{2\pi} \left(\partial_x \partial_y A_1 + \partial_z A_2 \right) \\
J^0 &= \frac{N}{2\pi} \left(\partial_x \partial_y B_1 + \partial_z B_2 \right).
\end{split}
\end{align}
It helps to think of the lattice operators in \figref{fig:ZNanisoH} as discretized versions of the above current densities,
  which can be made more precise by the correspondence in \eqnref{eq:operatorsAsFields}.

The Lagrangian is
\begin{align}
\begin{split}
L &= \frac{N}{2\pi} (A_1 \partial_t B_2 + A_2 \partial_t B_1)
    + B_0 \underbrace{\frac{N}{2\pi} \left(\partial_x \partial_y A_1 + \partial_z A_2 \right)}_{I^0}
   \\
 &\quad + A_0 \underbrace{\frac{N}{2\pi} \left(\partial_x \partial_y B_1
   + \partial_z B_2 \right)}_{J^0} - \sum_{a=0,1,2} (A_a J^a + B_a I^a).
\end{split}
\end{align}
There are six fields in total:
  $A_a$ and $B_a$ for $a=0,1,2$.
The first term results because $A$ and $B$ are conjugate fields.
In the next two terms,
  $B_0$ and $A_0$ act as Lagrange multipliers,
  which project into the ground state Hilbert space by projecting out excitations.
The final term couples $A$ and $B$ to source fields $J$ and $A$.
Similar to the lattice model,
  the field theory also exhibits a self-duality given by
\begin{align}
\begin{split}
  A_a &\leftrightarrow B_a \\
  J_a &\leftrightarrow I_a. \label{eq:QFT duality}
\end{split}
\end{align}

By construction, and the fact that the Hamiltonian terms commute, the Lagrangian exhibits a gauge invariance due to the vanishing Poisson bracket
\begin{equation}
  \{I^0(t,\vx),J^0(t,\vx')\} = 0
\end{equation}
where
\begin{align}
\begin{split}
  \{A_1(t,\vx),B_2(t,\vx')\} &= \{A_2(t,\vx),B_1(t,\vx')\} = \frac{2\pi}{N} \delta^3(\vx - \vx') \\
  \{A_1(t,\vx),B_1(t,\vx')\} &= \{A_1(t,\vx),B_2(t,\vx')\} = 0.
\end{split}
\end{align}
The vanishing Poisson bracket is the field theory analog of the fact that the terms in the lattice model commute.
The gauge transformation for $A_1$ and $A_2$ can be derived from
\begin{align}
\begin{split}
\forall_{a=1,2}: A_a(t,\vx) \to& A_a(t,\vx)
  - \int_{\vx'} \bigg\{ A_a(t,\vx), \\
  &\quad \underbrace{\frac{N}{2\pi} \left[\partial'_x \partial'_y B_1(t,\vx') + \partial'_z B_2(t,\vx') \right]}_{J^0(t,\vx')} \bigg\} \zeta(t,\vx').
\end{split}
\end{align}
The transformation for $A_0$ is then found by requiring that the Lagrangian is invariant under the transformation (ignoring the source field $J$ for now).
We then find that the Lagrangian is invariant under the following gauge transformation:
\begin{align}
\begin{split}
  A_0 &\to A_0 - \partial_t \zeta \\
  A_1 &\to A_1 + \partial_z \zeta \\
  A_2 &\to A_2 + \partial_x \partial_y \zeta. \label{eq:gaugeTrans}
\end{split}
\end{align}
As required by the duality (\eqnref{eq:QFT duality}),
  the Lagrangian is also invariant by a similar transformation of the $B$ field.
In order for the Lagrangian to be gauge invariant in the presence of the source fields,
  the source fields must obey the following conserved current constraints:
\begin{align}
\begin{split}
  \partial_t J^0 - \partial_z J^1 - \partial_x \partial_y J^2 &= 0 \\
  \partial_t I^0 - \partial_z I^1 - \partial_x \partial_y I^2 &= 0.
\end{split}
\end{align}

The field theory is invariant under the following form of spacetime transformations:
\begin{align}
  t &\to \tT(t) &
  x &\to \tX(x) &
  y &\to \tY(y) &
  z &\to \tZ(z)
\end{align}
where $\tilde{t}(t)$, $\tilde{x}(x)$, $\tilde{y}(y)$, and $\tilde{z}(z)$ are smooth and monotonic functions.
The gauge fields transform under the spacetime transformation as
\begin{align}
\begin{split}
  A_0(\R) &\to \frac{d\tT}{dt} A_0(\tR) \\
  A_1(\R) &\to \frac{d\tZ}{dz} A_1(\tR) \\
  A_2(\R) &\to \frac{d\tX}{dx} \frac{d\tY}{dy} A_2(\tR).
\end{split}
\end{align}

\end{document}